\documentclass{article}
\usepackage{mathtools,amsfonts,amssymb}
\usepackage[utf8]{inputenc}
\usepackage[no-natbib-sort]{jheppub}
\usepackage{todonotes}
\usepackage{hyperref}
\usepackage{bm}
\usepackage{booktabs}
\usepackage[nameinlink,capitalize]{cleveref}

\makeatletter{}
\g@addto@macro\bfseries{\boldmath}
\renewcommand{\@email}[1]{\texttt{#1}}
\makeatother

\newcommand*{\seccoord}[1]{\ensuremath\hat{#1}} 
\newcommand*{\secx}{\ensuremath\seccoord{x}} 
\newcommand*{\secy}{\ensuremath\seccoord{y}} 
\newcommand*{\secz}{\ensuremath\seccoord{z}} 

\newcommand*{\sechomol}[1]{\ensuremath\mathcal{#1}}
\newcommand*{\zerohomol}{\ensuremath\sechomol{Z}}
\newcommand*{\chargeonelocus}{\ensuremath V_{q = 1}}
\newcommand*{\sutwofundlocus}{\ensuremath\Delta_{(a)}}

\DeclareMathOperator{\SO}{SO}
\DeclareMathOperator{\SU}{SU}
\DeclareMathOperator{\U}{U}
\DeclareMathOperator{\gE}{E}

\DeclareMathOperator{\Res}{Res}
\newcommand*{\asu}{\ensuremath\mathfrak{su}}
\newcommand*{\au}{\ensuremath\mathfrak{u}}
\newcommand*{\cN}{\mathcal{N}}
\newcommand*{\cO}{\mathcal{O}}
\newcommand*{\cG}{\mathcal{G}}
\newcommand*{\cS}{\mathcal{S}}
\newcommand*{\F}{\mathbb{F}}
\newcommand*{\bP}{\mathbb{P}}
\newcommand*{\Z}{\mathbb{Z}}
\newcommand*{\C}{\mathbb{C}}

\newcommand*{\SMn}{\ensuremath\SU(3) \times \SU(2) \times \U(1)}
\newcommand*{\SM}{\ensuremath(\SU(3) \times \SU(2) \times \U(1)) / \Z_6}
\newcommand*{\sm}{\ensuremath\asu(3) \oplus \asu(2) \oplus \au(1)}
\newcommand*{\SMuh}{\ensuremath\SU(4) \times \SU(3) \times \SU(2)}
\newcommand*{\PS}{\ensuremath(\SU(4) \times \SU(2) \times \SU(2)) / \Z_2}

\newcommand{\dzerotilde}[0]{\ensuremath\tilde{d}_0}

\newcommand{\canonclass}[0]{\ensuremath K_B}

\newcommand*{\mat}[2][b]{\ensuremath\begin{#1matrix}#2\end{#1matrix}}

\newcommand{\locus}[1]{\ensuremath\{#1\}}
\newcommand{\tuning}[2]{\ensuremath#1 \to #2}

\title{General F-theory models with tuned $\SM$ symmetry}
\author[1]{Nikhil Raghuram,}
\author[2]{Washington Taylor,}
\author[2]{and Andrew P. Turner}

\affiliation[1]{Department of Physics \\
    Robeson Hall, 0435 \\
    Virginia Tech \\
    850 West Campus Drive \\
    Blacksburg, VA 24061, USA}
\affiliation[2]{Center for Theoretical Physics \\
    Department of Physics \\
    Massachusetts Institute of Technology \\
    77 Massachusetts Avenue \\
    Cambridge, MA 02139, USA}

\emailAdd{\texttt{nikhilr} \textrm{at} \texttt{vt.edu}}
\emailAdd{\texttt{wati} \textrm{at} \texttt{mit.edu}}
\emailAdd{\texttt{apturner} \textrm{at} \texttt{mit.edu}}

\preprint{MIT-CTP-5169}

\abstract{
We construct a general form for an F-theory Weierstrass model over a general
base giving a 6D or 4D supergravity theory with gauge group $\SM$ and generic
associated matter, which includes the matter content of the standard model.
The Weierstrass model is identified by unHiggsing a model with $\U(1)$ gauge
symmetry and charges $q \le 4$ previously found by the first author. This
model includes two distinct branches that were identified in earlier work, and
includes as a special case the class of models recently studied by Cveti\v{c},
Halverson, Lin, Liu, and Tian, for which we demonstrate explicitly the
possibility of unification through an $\SU(5)$ unHiggsing. We develop a
systematic methodology for checking that a parameterized class of F-theory
Weierstrass models with a given gauge group $G$ and fixed matter content is
generic (contains all allowed moduli) and confirm that this holds for the
models constructed here.
}

\begin{document}
\maketitle
\flushbottom

\section{Introduction}
\label{sec:intro}

A primary goal of string theory is to understand how the observed physics of
the standard model of particle physics can arise in a UV complete quantum
theory of gravity. Over the years, many different approaches have been taken
to realizing standard model-like physics in the context of string
compactifications, and recent work \cite{AndersonConstantinGrayLukasPalti,
ConstantinHeLukas, CveticEtAlQuadrillion} suggests that the number of such
possible realizations may be very large.  One of the features of F-theory
\cite{VafaF-theory,MorrisonVafaI,MorrisonVafaII} is that it gives a good
global picture of an enormous nonperturbative class of string
compactifications, so that one can begin to gain some insight into what
structures are typical and which require extensive fine tuning.

There are a number of different ways in which the gauge group of the standard
model could in principle arise in F-theory.  Some of these are reviewed in
\cref{sec:F-theory-SM}.  In this paper, we address the most straightforward
approach, in which the standard model gauge group is simply directly tuned in
the Weierstrass model describing the axiodilaton in the IIB/F-theory
framework. Such constructions of theories with the standard model gauge group
have been considered in \cite{LinWeigandSM,CveticEtAlThreeParam,
LinWeigandG4,CveticEtAlMSSMZ2}; recently, Cveti\v{c}, Halverson, Lin, Liu, and
Tian (CHLLT) \cite{CveticEtAlQuadrillion}, building on a toric construction
identified in \cite{KleversEtAlToric} and aspects of global group structure
studied in \cite{LawrieEtAlRational,GrimmKapferKleversArithmetic,CveticLinU1},
considered one class of such models that can be realized over any weak Fano
base, giving a large number of possible standard model-like constructions in
F-theory.  In the paper \cite{TaylorTurner321}, two of the authors of this
paper described what seems to be the most generic class of tuned F-theory
models that give the gauge group $\SM$; these models include the CHLLT models
as a particular subclass.  As reviewed further in \cref{sec:generic}, we focus
on constructions with the group $\SM$, since there is a well-defined sense in
which the generic matter content of models with the gauge group $\SMn$ does
not match the observed matter in the standard model
\cite{TaylorTurnerGeneric}.  A limitation of \cite{TaylorTurner321}, however,
is that the description given there of the generic $\SM$ model was somewhat
indirect.  In this paper, we make this class of models more concrete by giving
an explicit general form of the Weierstrass model that covers much of the
range of models that were identified through more indirect means in
\cite{TaylorTurner321}. This explicit Weierstrass realization also reveals
some cases where there is an obstruction to the F-theory realization despite
6D anomaly cancellation, as we discuss in more detail in \cref{sec:p2}.

The structure of this paper is as follows: in \cref{sec:background} we review
various aspects of previous work, including a rigorous definition of the
notion of generic used here, and introduce the notion of a generic
parameterized F-theory model with fixed gauge group $G$. In
\cref{sec:construction} we give the explicit construction of the Weierstrass
model for the generic F-theory model with gauge group $\SM$.  In the
subsequent two sections, \cref{sec:class-A,sec:class-B}, we examine in more
detail two distinct subclasses of these models and their unHiggsings,
connecting to the analysis of \cite{TaylorTurner321}.  In \cref{sec:matter},
we describe the matter content of these models in more detail.  In
\cref{sec:dimension}, we introduce a simple numerical technique to confirm
that our model is generic in the sense that it captures all dimensions of
moduli space for corresponding 6D supergravity theories. In \cref{sec:p2}, we
analyze the generic $\SM$ F-theory models over the base $\bP^2$. In
\cref{sec:morrison-park}, we show that our model can be realized as a
specialization of the generic Morrison--Park $\U(1)$ model
\cite{MorrisonParkU1}. In \cref{sec:bases}, we make some observations
regarding the range of geometries that support these constructions, and some
concluding remarks are made in \cref{sec:conclusions}.

\section{Some background}
\label{sec:background}

We begin with a brief review of generic matter in F-theory, some discussion
of the different ways in which the standard model may be realizable in
F-theory, and a review of the results of \cite{TaylorTurner321}.

\subsection{Weierstrass models and gauge groups in F-theory}
\label{sec:weierstrass}

For a general introduction to F-theory, see \cite{WeigandTASI}. An F-theory
compactification is associated with a Weierstrass model
\begin{equation}
y^2 = x^3 + f x + g\,,
\end{equation}
where $f, g$ are functions on a base manifold $B$ (more technically, $f, g$
are sections of the line bundles $\cO(-4 \canonclass), \cO(-6 \canonclass)$,
with $\canonclass$ the canonical class of $B$). F-theory can be thought of as
defining a nonperturbative compactification of type IIB string theory; when
$B$ is a complex surface this gives a 6D supergravity theory, and when $B$ is
a complex threefold this gives a 4D theory.  Note that, in general, $B$ is not
a Calabi--Yau manifold, but rather has a positive (i.e., effective)
anticanonical class $-\canonclass$. The Weierstrass model over $B$ defines an
elliptically fibered Calabi--Yau manifold over $B$, where the axiodilaton of
type IIB theory corresponds to the elliptic curve parameter $\tau$ defined by
the Weierstrass model over each point in the base.

Nonabelian gauge group factors are associated with codimension-one loci in the
base where the elliptic fiber degenerates.  Using the Kodaira classification,
the gauge algebra can be identified by the orders of vanishing of $f$, $g$,
and the discriminant locus $\Delta = 4 f^3 + 27 g^2$.  In general, there are
two ways in which such a Kodaira singularity can arise.  Over simple bases
like $\bP^2$ or $\bP^3$ (or any other weak Fano base), the generic Weierstrass
model has no codimension-one singularities and the gauge group is trivial.
Over such bases, a gauge group can be ``tuned'' by restricting the form of the
Weierstrass model to ensure a certain type of Kodaira singularity.  Over bases
that are not weak Fano, for example the 2D Hirzebruch surfaces $\F_m$ with $m
\ge 3$, the anticanonical class $-\canonclass$ generally contains rigid
components over which there are ``non-Higgsable'' gauge group factors
\cite{MorrisonTaylorClusters,MorrisonTaylor4DClusters}.

The
gauge group can have additional $\U(1)$ factors when the elliptic fibration
admits extra rational sections. According to the Mordell--Weil theorem, the
sections of an elliptic fibration form the finitely generated group $\Z^r
\oplus \cG$ under elliptic curve addition, where $\cG$ is some
finite group \cite{LangNeron}. In the most basic situation, where the only
rational section of the elliptic fibration is the zero section, $\cG$ is
trivial, and $r$, an integer known as the Mordell--Weil rank, is 0. But we can
have non-trivial $\cG$ and $r$ when there are sections other than the zero
section. The finite part $\cG$ is generated by torsional sections, which have
finite order under elliptic curve addition. The $\Z^r$ subgroup, meanwhile, is
generated by $r$ sections of infinite order. When an elliptic fibration
describing an F-theory model has a non-trivial Mordell--Weil rank $r$, the
resulting gauge group includes a $\U(1)^r$ gauge factor \cite{MorrisonVafaII}.
In other words, extra rational sections (of infinite order) signal the
presence of additional $\U(1)$ factors in the gauge group. Importantly, the
resulting $\U(1)$ factors are not associated with a codimension-one locus in
the base with $\U(1)$ factors; they are in some sense a global feature of the
model. In this work, all of the $\U(1)$ gauge factors arise due to the
presence of additional rational sections.

On a practical level, rational sections occur when there are solutions for $x$
and $y$ in the Weierstrass equation that are, at least informally, rational
functions of the base coordinates. It is often easier to work with the global
Weierstrass form
\begin{equation}
y^2 = x^3 + f x z^4 + g z^6\,,
\end{equation}
where $[x:y:z]$ are the homogeneous coordinates of $\bP^{2, 3, 1}$. Sections
are then described by expressions $[\secx : \secy : \secz]$ written in terms of
the base coordinates that solve the global Weierstrass equation. We can always
recover the more typical Weierstrass form above from the global form by
setting $z$ to 1, which may seem to make the global Weierstrass form
superfluous. However, the global form offers a few advantages. The zero
section can be more transparently written as $[1 : 1 : 0]$ when using the global
Weierstrass form. Moreover, we can use the $[x : y : z] \to [\lambda^2
x : \lambda^3 y : \lambda z]$ rescaling to remove denominators from the $\secx$
and $\secy$ components of the rational sections. As a result, the sections can
be described in a more convenient fashion, and we therefore make use of the global
Weierstrass form at various points in this work.

\subsection{Generic matter}
\label{sec:generic}

The appearance of (geometrically) non-Higgsable gauge factors in F-theory
models over bases that are not weak Fano has the physical consequence that, in
many branches of F-theory, there are gauge groups that are ``generic'' in the
sense that they are present everywhere in that branch of the geometric moduli
space. For six-dimensional theories, there is a direct correspondence between
the geometric and physical moduli spaces, so that a branch of the theory with
a non-Higgsable gauge group corresponds to a component of the moduli space
where the gauge group arises everywhere.  In four dimensions, the story is
complicated by the presence of fluxes and the superpotential.

There is also a notion of genericity associated with certain matter
representations that can be made rigorous in six dimensions, and which carries
over naturally to F-theory compactifications in four dimensions.  As described
in \cite{TaylorTurnerGeneric}, in a 6D supergravity theory with a fixed
(tuned) gauge group $G$ and fixed (and relatively small) anomaly coefficients,
the set of generic matter representations corresponds to those matter fields
that are found on the branch of the moduli space with largest dimension.
Note that this definition of generic is well-defined in 6D supergravity
without reference to F-theory or any other UV completion.  Note also that
while the matter representations are generic for a chosen gauge group $G$, in
general the gauge group $G$ itself will not be a generic feature on that
branch of the moduli space, i.e., it can be broken through the Higgs mechanism
by giving expectation values to some of the matter fields.

As simple examples, for a $\U(1)$ gauge theory in 6D supergravity the generic
charged matter content contains only matter with charges $q = \pm1, \pm2$, and
for an $\SU(N)$ gauge theory the generic matter content consists of the
fundamental, adjoint, and two-index antisymmetric representations (with the
last of these only included when $N > 3$). On the other hand, $\U(1)$ charges
$q =\pm3$ or greater, or the two-index symmetric representation of $\SU(N)$
for $N > 2$, for example, are non-generic (``exotic'') matter representations
in this sense. Note that for algebras like $\asu(N) \oplus \au(1)$, the
generic matter depends on the global structure of the gauge group. For
example, when the structure is $(\SU(N) \times \U(1)) / \Z_N$, the $\U(1)$
charge is naturally measured in units of $1 / N$ and for the simplest
embedding the jointly charged generic matter representations are $\bm{N}_{1 /
N}$ and $\bm{N}_{1 / N \pm 1}$ (see for example \cite{LawrieEtAlRational,
GrimmKapferKleversArithmetic,CveticLinU1,TaylorTurnerGeneric} for discussion
of such issues).

This notion of generic matter for a given $G$ matches well with both anomaly
cancellation in 6D and with the framework of F-theory.  For the simplest
groups $G$, in particular those without many $\U(1)$ factors, the number of
generic matter representations matches with the number of anomaly cancellation
conditions, so the generic matter spectrum is essentially uniquely determined
by the gauge group and anomaly coefficients.  This relation becomes more
complicated particularly as the number of $\U(1)$ factors increases, where
there are different combinations of charges compatible with the generic matter
definition.  Also, for the simplest groups it turns out that the generic
matter types are precisely those realized by the simplest and least singular
F-theory constructions, such as for example those described in
\cite{BershadskyEtAlSingularities}.  In this way, and also by considering the
dimension of the underlying geometric moduli space, the notion of generic
matter naturally generalizes to 4D F-theory models.

\subsection{Generic parameterized F-theory models with fixed $G$}
\label{sec:generic-f-models}

Given the notion of generic matter, a natural question in F-theory is whether
a construction can be found for the most general model with a given gauge
group $G$ and the associated generic matter content.  Such a construction
would be realized by a class of Weierstrass models parameterized by various
sections of certain line bundles over the base, which realize the desired
gauge group.\footnote{Note that the parameters in these models, associated
with sections of certain line bundles, can be understood as divisors on the
F-theory base manifold, which in general represent multiple independent
complex degrees of freedom associated with the number of linearly independent
sections of the line bundle.  In a local or toric representation we can
understand these complex degrees of freedom in terms of the coefficients of
independent monomials in a local coordinate representation of the section.} We
refer to a parameterized class of Weierstrass models as ``generic $G$ F-theory
models'' when the following conditions are satisfied: first, the models in the
class should realize F-theory constructions with gauge group $G$ and the
associated generic matter; second, the model should be general in the sense
that it realizes the full connected moduli space of such models over any given
base.  In \cref{sec:dimension}, we explore the second of these conditions for
some known models as well as the new class constructed in this paper, and
provide a simple algorithm for checking if a known parameterized Weierstrass
model satisfies the second condition.  As an example, the Morrison--Park
$\U(1)$ model \cite{MorrisonParkU1} is confirmed to be the generic $\U(1)$
F-theory model in the sense described here.

As observed in \cite{TaylorTurnerGeneric}, the matter content of the standard
model is far from generic for the gauge group $\SMn$.  In particular, generic
matter for this gauge group contains no fields that are charged under all
three gauge group factors.  On the other hand, for the gauge group $\SM$, the
standard model fields are contained in the set of generic matter
representations.  Thus, in looking for tuned F-theory constructions of the
standard model, we look for the generic F-theory model with gauge group $\SM$.
Constructing an explicit Weierstrass model for this class of generic F-theory
models is the principal goal of this paper.  A more general question is how to
construct generic $G$ F-theory models for more general gauge groups with a
product of nonabelian and abelian $\U(1)$ factors.

\subsection{Approaches to the standard model in F-theory}
\label{sec:F-theory-SM}

To put this work in context, it may be helpful to briefly summarize some of
the different ways that the standard model may be realized in F-theory.  A
table of possibilities is shown in \cref{fig:SM-in-F-theory}.  In this table,
the rows correspond to whether the gauge group is realized through tuning or
non-Higgsable structure (genericity), and the columns correspond to whether or
not there is a (geometric) unified group broken by fluxes.  We make some brief
comments on each of these possible scenarios.

\begin{table}
\centering

\renewcommand{\arraystretch}{1.5}
\begin{tabular}{r|c|c|} \cline{2-3}
& GUT & $\SMn$ \\ \hline
\multicolumn{1}{|c|}{Tuned $G$} & Tuned GUT (e.g., $\SU(5)$)
    & Directly tuned $G_\text{SM}$ \\ \hline
\multicolumn{1}{|c|}{Non-Higgsable $G$} & Non-Higgsable GUT (e.g., $\gE_6$)
	& Non-Higgsable $G_\text{SM}$ \\ \hline
\end{tabular}

\caption{Different approaches to realizing the standard model in F-theory can
involve combining either tuning or non-Higgsable structure with either
unification (GUTs) or non-unified standard model gauge group structure}
\label{fig:SM-in-F-theory}
\end{table}

\paragraph{Tuned GUT scenarios:}
This is the first approach that was explored to F-theory realizations of the
standard model, and the one that has been studied by far the most in the
literature.  The general idea is to start with a Weierstrass model with a
tuned unification group such as $\SU(5)$ and to use fluxes to break the
$\SU(5)$ down to the standard model.  This approach was initiated in
\cite{DonagiWijnholtGUTs,DonagiWijnholtModelBuilding,BeasleyHeckmanVafaI,
BeasleyHeckmanVafaII}; for an overview of work in this direction see
\cite{HeckmanReview,WeigandLectures,MaharanaPaltiReview,WeigandTASI}.

\paragraph{SM with non-Higgsable group scenarios:}
In this approach, one could use non-Higgsable (geometrically generic)
structure for at least the nonabelian $\SU(3) \times \SU(2)$ part of the
standard model gauge group.  This approach was explored in
\cite{GrassiHalversonShanesonTaylor}, and is possible since the group factors
$\SU(3) \times \SU(2)$ can appear as non-Higgsable factors with jointly
charged matter in 4D (but not 6D) F-theory models.  One could also try to
incorporate a non-Higgsable $\U(1)$ factor, like those identified in
\cite{MartiniTaylorSemitoric, WangU1s}, but this requires very specific base
structure and is difficult to make compatible with the necessary nonabelian
gauge factors in a geometrically generic fashion.

\paragraph{Non-Higgsable GUT scenarios:}
In this class of scenarios, one starts with a non-Higgsable unification group
such as $\gE_6$, $\gE_7$, or $\gE_8$ and then carries out flux breaking to the
standard model.  Since the vast majority of allowed threefold bases give rise
to non-Higgsable $E_6, E_7,$ or $E_8$ factors \cite{TaylorWangMC,
HalversonLongSungAlg, TaylorWangLandscape}, this approach seems feasible over
the widest range of bases and does not involve fine tuning of the Weierstrass
model.  Indeed, a naive estimate of flux vacua suggests that the set of flux
vacua may be dominated by a certain elliptic Calabi--Yau fourfold geometry
\cite{TaylorWangVacua}, in which the only possible approach to realizing the
standard model seems to be through a non-Higgsable $\gE_8$.  While it seems
difficult to realize the standard model Yukawa couplings in this geometry due
to general arguments given in \cite{BeasleyHeckmanVafaI}, it may be possible
to get around this by realizing the standard model matter through SCFT sectors
arising as $\gE_8$ conformal matter \cite{TianWangEString}.

\paragraph{Directly tuned scenarios:}
This is the approach we consider in this paper, in which the full gauge group
is tuned in the Weierstrass model.  As discussed in the introduction, models
of this type were previously considered in
\cite{LinWeigandSM,CveticEtAlThreeParam,LinWeigandG4,CveticEtAlMSSMZ2,
CveticEtAlQuadrillion,TaylorTurner321}. While these models can largely arise
from deformations (which always can be understood in terms of Higgsing
processes in the 6D context) from larger groups, the breaking in these cases
of any possible GUT depends upon deformations of the Weierstrass model.
\\[-0.3\baselineskip]

In addition to these basic different types of constructions associated with
different kinds of Weierstrass models, there are other possibilities.  For
example, part of the gauge group may come from D3-branes in the type II
context and not from D7-branes.\footnote{Thanks to Yinan Wang for discussions
on this point.}  It is an interesting question to consider the relative
features of these different constructions, both in terms of matching
phenomenologically observed aspects of the standard model and in terms of
relative frequency in the broad F-theory landscape.  While a naive application
of standard flux counting arguments
\cite{DouglasStatistics,AshokDouglas,DenefDouglas,DenefLesHouches} might
suggest that most compactifications are associated with the geometry analyzed
in \cite{TaylorWangVacua}, a more complete analysis including geometric
factors for the density of flux vacua may give an additional
exponential weighting to
bases giving Calabi--Yau fourfolds with smaller $h^{3, 1}$
\cite{ChengMoorePaquette}, which would suggest that typical 4D F-theory vacua
may be associated with Calabi--Yau fourfolds that have more room for tuning
gauge factors beyond the non-Higgsable structure.  A simple counting of the
number of Weierstrass parameters that must be tuned to realize $\SU(5)$ GUTs
over simple bases suggests that these vacua involve much fine tuning and may
be statistically disfavored \cite{BraunWatariGenerations}; similar arguments
apply to tuned standard model gauge group constructions like those considered
here though the tuning is less extreme.  On the other hand, it may be that
over many bases, and in the presence of fluxes, the number of parameters that
needs to be tuned may become much smaller; in fact, fluxes may force the
geometry to loci where certain groups are more prevalent.  Thus, there is no
obvious argument that any of these four possible scenarios is completely ruled
out or necessarily overwhelmingly dominant in the F-theory landscape of $\cN =
1$ 4D string vacua. A pragmatic approach at this time is to consider all the
possibilities, exploring each to the extent possible, while simultaneously
trying to understand better the statistical distribution of the different
types of associated vacua in the landscape and the role of fluxes and the
superpotential in pushing the geometry to certain loci or breaking
non-Higgsable GUT groups to structures like the standard model.

\subsection{Review of previous work on generic $\SM$ models}
\label{sec:review}

We now summarize the results of \cite{TaylorTurner321} and review notation
introduced there.

We first consider generic $\SM$ models in 6D supergravity. As discussed in
\cite{TaylorTurnerGeneric}, there are ten generic charged matter fields and
ten nontrivial 6D anomaly cancellation conditions constraining charged matter
for the gauge group $\SM$. Thus, the anomaly cancellation conditions can
generally be solved exactly to give the multiplicities of generic matter
representations, given a fixed choice of anomaly coefficients $a, b_3, b_2,
\tilde{b}$, associated respectively with gravity and the gauge factors
$\SU(3)$, $\SU(2)$, and $\U(1)$. It is convenient to define the quantities
$\beta, X, Y$ given by
\begin{equation}
\label{eq:betaXY}
\begin{aligned}
\tilde{b} &= \frac{4}{3} b_3 + \frac{3}{2} b_2 + 2 \beta\,, \\
X &= -8 a - 4 b_3 - 3 b_2 - 2 \beta\,, \\
Y &= a + b_3 + b_2 + \beta\,.
\end{aligned}
\end{equation}
Solving the anomaly cancellation conditions yields the matter multiplicities
given in \cref{tab:generic}.

\begin{table}[h!]
\centering

\[\begin{array}{ccc}\toprule
\text{Generic Matter} & \text{Multiplicity} & \text{MSSM Multiplet} \\ \midrule
\left(\bm{3}, \bm{2}\right)_{\mathrlap{1 / 6}} & b_3 \cdot b_2 &
    \bm{Q} \\[0.8em]
\left(\bm{3}, \bm{1}\right)_{\mathrlap{-4 / 3}} & b_3 \cdot Y & \\[0.8em]
\left(\bm{3}, \bm{1}\right)_{\mathrlap{-1 / 3}} & b_3 \cdot X &
    \overline{\bm{D}}^c \\[0.8em]
\left(\bm{3}, \bm{1}\right)_{\mathrlap{2 / 3}} & b_3 \cdot (\beta - 2 a) &
    \overline{\bm{U}}^c \\[0.8em]
\left(\bm{1}, \bm{2}\right)_{\mathrlap{1 / 2}} & b_2 \cdot (X + \beta - a) &
    \overline{\bm{L}}, \bm{H}_u, \overline{\bm{H}_d}  \\[0.8em]
\left(\bm{1}, \bm{2}\right)_{\mathrlap{3 / 2}} & b_2 \cdot Y & \\[0.8em]
(\bm{1}, \bm{1})_{\mathrlap{1}} & (b_3+ b_2+2 \beta) \cdot X - a \cdot b_2 &
    \bm{E}^c \\[0.8em]
(\bm{1}, \bm{1})_{\mathrlap{2}} & \beta \cdot Y & \\[0.8em]
(\bm{8}, \bm{1})_{\mathrlap{0}} & 1 + b_3 \cdot (b_3 + a) / 2 & \\[0.8em]
(\bm{1}, \bm{3})_{\mathrlap{0}} & 1 + b_2 \cdot (b_2 + a) / 2 & \\ \bottomrule
\end{array}\]

\caption{Generic matter representations (not including conjugates) charged
under the gauge group $\SM$, along with multiplicities for the generic matter
solution of the 6D anomaly equations. This includes all the charged MSSM
multiplets. The parameters $\beta, X, Y$ are defined in \cref{eq:betaXY}.}
\label{tab:generic}
\end{table}

We then proceed to classify the anomaly-consistent models using intuition
gained in the case of no tensor multiplets ($T = 0$), for which the anomaly
coefficients $a, b_3, b_2, \tilde{b}$ are simply integers (in general, they
are vectors in an $\SO(1, T)$ lattice). In order to have nontrivial nonabelian
gauge factors with properly-signed kinetic terms, we must have $b_3, b_2 > 0$.
Given that we require the spectra to have non-negative multiplicities, this
immediately implies that $X, Y \ge 0$. We then consider two non-disjoint
classes of solutions, which together cover all 6D $T = 0$ models:
\begin{quote}
\begin{description}
\item[Class (A)] models with $\beta \ge 0$,
\item[Class (B)] models with $Y = 0$ (for which we may have $\beta < 0$).
\end{description}
\end{quote}
Although these two classes were defined using intuition gained in the case of
6D $T = 0$, we can generalize them to arbitrary numbers of tensor multiplets,
although for $T > 0$ it is not guaranteed that every anomaly-consistent
solution falls into one or both of these classes.

As discussed in \cite{TaylorTurner321}, these two classes of models appear to
have good constructions in F-theory that generalize naturally to 4D
supergravity theories. Specifically, the $\beta \ge 0$ (Class (A)) models have
spectra consistent with a Higgsing deformation of an $\SMuh$ model with
respective gauge anomaly coefficients $B_4 = b_3, B_3 = b_2, B_2 = \beta$, and
we would thus expect that many of these models could be constructed in
F-theory by starting with a Tate-type tuning
\cite{BershadskyEtAlSingularities, KatzEtAlTate} of $\SMuh$ (see
\cref{sec:vanishing-discriminant}) and carrying out the deformation in the
Weierstrass model.  As we discuss in \cref{sec:p2}, however, such a Tate
construction is not possible when $b_3, b_2, \beta$ are too large, even when
anomaly cancellation naively suggests that such a model should exist. The $Y =
0$ models correspond to (a slight generalization of) the F-theory models with
toric fiber $F_{11}$ discussed in \cite{KleversEtAlToric}; the multiplicities
given there are matched to those given in \cref{tab:generic} by the
identification $b_3 = \cS_9, b_2 = \cS_7 - \cS_9 - \canonclass$. Additionally,
most of the models in the $Y = 0$ class (Class (B)) admit an unHiggsing to a
Pati--Salam $\PS$ model with respective gauge anomaly coefficients $B_4 = b_3,
B_2 = b_2, B_2^\prime = -4 a - 2 b_3 - b_2$, and we would expect that these
models can be realized in F-theory by a Tate tuning of this gauge group
followed by a deformation of the Weierstrass model. The $Y = 0$ model with
$b_3 = b_2 = -a$ has not only a toric fiber $F_{11}$ description and a
Pati--Salam description, but also has a spectrum compatible with an unHiggsing
to $\SU(5)$ with gauge anomaly coefficient $B_5 = -a$.

Thus, we expect that we can construct many, and perhaps all, of the models in
the above two classes in F-theory.  The models of Class (A) are parameterized
by three divisor classes $b_3, b_2, \beta$, which can be varied independently
subject to the constraints that the multiplicities in \cref{tab:generic} are
non-negative, while the models of class (B) have the additional constraint $Y
= 0$ and thus are parameterized by only the two independent divisor classes
$b_3, b_2$.  Even in cases where the Tate tuning of the unHiggsed $\SU(4)
\times \SU(3) \times \SU(2)$ or Pati--Salam model is possible, however, this
procedure is difficult to make explicit since it is hard to identify the
Higgsing deformations of the given nonabelian Weierstrass model. In the
current paper, we give an explicit Weierstrass model that realizes these
models in all cases where the Tate tunings of the enhanced gauge groups
discussed above yield the desired gauge algebra; as we discuss in
\cref{sec:p2}, there are cases where there is accidental enhancement of the
gauge algebra beyond $\sm$.

\section{Obtaining the model}
\label{sec:construction}

In this and the following two sections, we give a direct, explicit description
of the desired $\SM$ Weierstrass models that is relevant for both 6D and 4D
constructions.

\subsection{Weierstrass model}
\label{sec:weierstrass-model}

We start with a Weierstrass model, first described in
\cite{Raghuram34}, that supports a $\U(1)$ gauge group and matter with
charges $\pm1$ through $\pm4$.  This Weierstrass model is rather
lengthy, so we will not write it down here.  The model can be enhanced
to support a larger gauge group by setting the divisor parameters $a_1$ and
$s_3$ to zero, leading to an $f$ and $g$ of the form
\begin{equation}
\label{eq:su321weierstrass}
\begin{aligned}
f &= -\frac{1}{48} \left[s_6^2 - 4 b_1 (d_0 s_5 + d_1 s_2)\right]^2 \\
&\quad + \frac{1}{2} b_1 d_0 \left[2 b_1 \left(d_0 s_1 s_8 + d_1 s_2 s_5 + d_2 s_2^2\right) - s_6 (s_2 s_8 + b_1 d_1 s_1)\right]\,, \\
g &= \frac{1}{864} \left[s_6^2 - 4 b_1 (d_0 s_5 + d_1 s_2)\right]^3 + \frac{1}{4} b_1^2 d_0^2 \left(s_2 s_8 - b_1 d_1 s_1\right)^2 - b_1^3 d_0^2 d_2 \left(s_2^2 s_5 - s_2 s_1 s_6 + b_1 d_0 s_1^2\right) \\
&\quad - \frac{1}{24} b_1 d_0 \left[s_6^2 - 4 b_1 (d_0 s_5 + d_1 s_2)\right] \left[2 b_1 \left(d_0 s_1 s_8 + d_1 s_2 s_5 + d_2 s_2^2\right) - s_6 (s_2 s_8 + b_1 d_1 s_1)\right]\,.
\end{aligned}
\end{equation}
The homology classes for the parameters are given in
the first column of \cref{tab:su321homology}.

\begin{table}
\centering

$\begin{array}{ccc}\toprule
\text{Parameter} & b_3, b_2, \beta \text{ Basis} & b_3, b_2, Y \text{ Basis} \\ \midrule
b_1 & b_3 & b_3 \\
d_0 & b_2 & b_2 \\
s_1 & \beta & -\canonclass + Y - b_3 - b_2 \\
s_2 & \canonclass + b_3 + b_2 + \beta & Y \\
d_1 & -3 \canonclass - 2 b_3 - b_2 - \beta & -2 \canonclass - Y - b_3 \\
d_2 & -6 \canonclass - 4 b_3 - 3 b_2 - 2 \beta & -4 \canonclass - 2 Y - 2 b_3 - b_2 \\
s_5 & -2 \canonclass - b_3 - b_2 & -2 \canonclass - b_3 - b_2 \\
s_6 & -\canonclass & -\canonclass \\
s_8 & -4 \canonclass - 2 b_3 - 2 b_2 - \beta & -3 \canonclass - Y - b_3 - b_2 \\ \bottomrule
\end{array}$

\caption{Homology classes for the divisors parameterizing the $\SM$
Weierstrass model in \cref{eq:su321weierstrass}, in two bases.}
\label{tab:su321homology}
\end{table}

The resulting discriminant is proportional to $b_1^3 d_0^2$, indicating the
presence of $I_2$ singularities along $\locus{d_0 = 0}$ and $I_3$
singularities along $\locus{b_1 = 0}$ . The $I_2$ singularities signal that
the model has an $\SU(2)$ gauge symmetry tuned on $\locus{d_0 = 0}$. One can
additionally verify that the split condition is satisfied for the $I_3$
singularities and that the model admits an $\SU(3)$ gauge symmetry tuned on
$\locus{b_1 = 0}$. But the model also has an extra non-torsional section with
components given by
\begin{equation}
\label{eq:su321section}
\begin{aligned}
\secx &= \left(b_1 d_0 s_1 - \frac{1}{2} s_2 s_6\right)^2 - \frac{1}{6} s_2^2 \left(s_6^2 + 2 b_1 d_1 s_2 - 4 b_1 d_0 s_5\right)\,, \\
\secy &= -\left(b_1 d_0 s_1 - \frac{1}{2} s_2 s_6\right)^3 + \frac{1}{4} s_2^2 \left(b_1 d_0 s_1 - \frac{1}{2} s_2 s_6\right) \left(s_6^2 + 2 b_1 d_1 s_2 - 4 b_1 d_0 s_5\right) \\
&\quad + \frac{1}{4} s_2^4 b_1 \left(d_1 s_6 - 2 d_0 s_8\right)\,, \\
\secz &= s_2\,.
\end{aligned}
\end{equation}
As a result, we have an additional $\U(1)$ gauge factor, and the total gauge
algebra is that of $\asu(3) \oplus \asu(2) \oplus \au(1)$, where the
global structure of the group may have a quotient by a discrete group in the
$\Z_6$ center.

It is helpful to calculate the height $\tilde{b}$ of the generating section
\labelcref{eq:su321section}, as it captures physical information about the
$\U(1)$ factor. In general, the height is given by the formula
\cite{ParkIntersection, MorrisonParkU1}
\begin{equation}
\tilde{b} = -2 \canonclass + 2 \, \pi(\sechomol{S} \cdot \zerohomol) - \left(\mathcal{R}^{-1}_\kappa\right)_{I J} \left(\sechomol{S} \cdot \alpha_{\kappa, I}\right) \left(\sechomol{S} \cdot \alpha_{\kappa, J}\right) b_\kappa\,,
\end{equation}
where $\sechomol{S}$ is the homology class of the generating section,
$\zerohomol$ is the homology class of the zero section, and $\alpha_{\kappa,
I}$ is the $I$th exceptional divisor associated with the $\kappa$th nonabelian
gauge factor. The map $\pi$ is the projection onto the base. Finally,
$b_\kappa$ is the divisor in the base supporting the $\kappa$th nonabelian
gauge factor, while $\mathcal{R}_\kappa$ is the normalized root matrix for the
gauge factor. For the case at hand, the generating section meets the zero
section whenever $s_2$ is zero, so $\pi(\sechomol{S} \cdot \zerohomol)$ is
equal to $[s_2] = Y$. Meanwhile, the $\secy$ component of the generating
section is proportional to both $b_1$ and $d_0$, so for both the $\SU(3)$ and
$\SU(2)$ gauge factors, $\left(\sechomol{S} \cdot \alpha_{\kappa, I}\right)$
is nonzero for at least one of the exceptional divisors. For the $\SU(2)$
factor, there is only one exceptional divisor, and $\mathcal{R}^{-1}_\kappa$
is essentially the constant $\frac{1}{2}$. For the $\SU(3)$ factor, the
generating section will hit one of the two exceptional divisors, which without
loss of generality we can take to be $I = 1$. Additionally,
$\left(\mathcal{R}^{-1}_\kappa\right)_{1 1}$ is $\frac{2}{3}$. Combining all
of this information together, we have
\begin{equation}
\tilde{b} = -2 \canonclass + 2 Y - \frac{1}{2} b_2 - \frac{2}{3} b_3 =
  \frac{4}{3} b_3 + \frac{3}{2} b_2 + 2 \beta\,.
\end{equation}
Note that this expression matches the first relation in \cref{eq:betaXY}.

We assert that \cref{eq:su321weierstrass} is in fact the desired generic
F-theory model with gauge group $\SM$.  In the succeeding sections we perform
a variety of computations that support this hypothesis.  In particular, we
show that a generic $\SM$ model can be Higgsed back to the $q = 3, 4$ model
that was the starting point of this construction, we show that the model
\labelcref{eq:su321weierstrass} exhibits the two classes of constructions
identified in \cite{TaylorTurner321} and can be unHiggsed to the associated
parent nonabelian groups in each case, we identify the matter loci of
\cref{eq:su321weierstrass} as appropriate for the generic $\SM$ model, and
finally show that the dimensionality of the parameterized Weierstrass model
\labelcref{eq:su321weierstrass} matches with that expected for 6D models, at
least in the cases with no tensor multiplets where the Weierstrass model gives
the desired gauge group.  Taken together these analyses demonstrate
definitively that the model \labelcref{eq:su321weierstrass} is indeed the
generic Weierstrass model for F-theory constructions  with gauge group $\SM$.
The correspondence between the parameters in the Weierstrass model
\labelcref{eq:su321weierstrass} and the parameters reviewed in
\cref{sec:review} is given in \cref{tab:su321homology}.

\subsection{Higgsing to $\U(1)$ with $q = 4$}
\label{sec:higgsing-to-1}

As the Weierstrass model \labelcref{eq:su321weierstrass} was found via an
unHiggsing of the charge-$4$ $\U(1)$ Weierstrass model, it is useful to
consider the corresponding Higgsing from the field theory point of view.
Specifically, we consider Higgsings of the gauge group $\SM$ induced by giving
nonzero VEVs to weights in the associated generic matter representations. Up
to Weyl reflection, there are $35$ distinct embeddings of $\U(1)$ into $\SM$
that can be reached by such a Higgsing. By comparing the matter multiplicities
in \cref{tab:codimtwoloci} with those in Table~7 of \cite{Raghuram34}, we find
that the relevant Higgsing leaves unbroken the $\U(1)$ generated by
\begin{equation}
\mu = \frac{4}{3} \lambda_3^{(1)} + \frac{8}{3} \lambda_3^{(2)} + \frac{3}{2} \lambda_2 + \lambda_1\,,
\end{equation}
where $\lambda_3^{(1)} = \mat[p]{1 & 0 & 0 \\ 0 & -1 & 0 \\ 0 & 0 & 0},
\lambda_3^{(2)} = \mat[p]{0 & 0 & 0 \\ 0 & 1 & 0 \\ 0 & 0 & -1}$ are the
$\SU(3)$ Cartan generators and $\lambda_2, \lambda_1$ are respectively the
$\SU(2), \U(1)$ Cartan generators. There are $24$ different choices of three
generic weights that can be given nonzero VEVs to yield this $\U(1)$ embedding
after Higgsing. By considering intermediate Higgsings that can be seen
explicitly in the Weierstrass model \labelcref{eq:su321weierstrass}, we
determine that, up to Weyl reflection, the relevant Higgsing in our case gives
nonzero VEVs to the weights $\left(1, 0, 0, -\frac{4}{3}\right), \left(-1, 1,
0, -\frac{4}{3}\right), \left(0, 0, -1, \frac{3}{2}\right)$, respectively in
the representations $(\bm{3}, \bm{1})_{-4 / 3}, (\bm{3},
\bm{1})_{-4 / 3}, (\bm{1}, \bm{2})_{3 / 2}$, at least when there
are enough fields to satisfy the D-term constraints.

As mentioned above, because three weights must be given VEVs to achieve this
Higgsing, at the group theoretic level there are intermediate gauge algebras
that can be reached by giving VEVs to a subset of these weights. Indeed, these
intermediate subalgebras can be realized as intermediate unHiggsings of the $q
= 4$ $\U(1)$ Weierstrass model. Specifically, the intermediate algebras
$\asu(3) \oplus \au(1)$, $\asu(2)^\prime \oplus \asu(2) \oplus \au(1)$,
$\asu(2) \oplus \au(1)$, and $\asu(2)^\prime \oplus \au(1)$ can be reached,
where the prime is used to indicate an $\asu(2)$ subalgebra of the $\asu(3)$
factor. For example, taking $\tuning{a_1}{a_1^\prime b_1}$ enhances the
$\au(1)$ algebra to $\asu(2)^\prime \oplus \au(1)$, and subsequently taking
$\tuning{a_1^\prime}{a_1^{\prime \prime} b_1}, \tuning{s_3}{s_3^\prime b_1}$
further enhances this to $\asu(3) \oplus \au(1)$. Similar enhancements can be
made to explicitly realize all of the above intermediate Higgsings of $\SM$ in
the Weierstrass model \labelcref{eq:su321weierstrass}.

\section{Class (A) models}
\label{sec:class-A}

As discussed in \cref{sec:review}, there is an important subclass of $\SM$
models, the Class (A) models, for which the divisor class $\beta$ is
effective. This presents the question of whether the Weierstrass model
\labelcref{eq:su321weierstrass} can realize these models. Because there is no
prior Weierstrass realization of the Class (A) models, this question has added
significance. There are some easily seen features of the Weierstrass model
that suggest that it can, in fact, support the Class (A) models. For instance,
our Weierstrass model has three independent divisor classes when all
parameters are non-vanishing, just as seen in the analysis of
\cite{TaylorTurner321}. And the requirement that $\beta$ is effective has a
concrete interpretation in the Weierstrass model, as it implies that we can
take the parameter $s_1$ in \cref{eq:su321weierstrass} to be nonzero. Thus,
Class (A) is in some sense the more general of the two classes.

A more thorough way of addressing this question is to examine how we can
further unHiggs the Weierstrass model \labelcref{eq:su321weierstrass}.  The
Class (A) models are expected from \cite{TaylorTurner321} to admit an
unHiggsing to $\SMuh$. In fact, the Weierstrass model generally allows for
such an unHiggsing. If we set $\tuning{s_2}{0}$, the discriminant then becomes
proportional to $b_1^4 d_0^3 s_1^2$, indicating that we have $I_4$ fibers
along $\locus{b_1 = 0}$, $I_3$ fibers along $\locus{d_0 = 0}$, and $I_2$
fibers along $\locus{s_1 = 0}$. The split condition is also satisfied for the
$\locus{b_1 = 0}$ and $\locus{d_0 = 0}$ singular loci. Therefore, the enhanced
gauge group is $\SMuh$, with the gauge group factors tuned on the exact
divisor classes predicted in \cite{TaylorTurner321}. The divisors $\locus{b_1
= 0}$, $\locus{d_0 = 0}$, and $\locus{s_1 = 0}$ do not have any singular
structure, so one would expect this model to have a generic matter spectrum.
An explicit matter analysis shows this to be the case, and the spectrum agrees
exactly with the expectations in \cite{TaylorTurner321} for the unHiggsed
model. Since we see an enhancement to $\SMuh$, our Weierstrass model should
realize the Class (A) spectra. We will see further confirmation of this fact
when we explicitly determine the matter spectrum of the $\SM$ model in
\cref{sec:matter}.

Before going on, it is worth understanding the enhancement to $\SMuh$ in more
detail. From field theoretic arguments, we know that $\SMuh$ can be Higgsed
down to $\SM$ by giving VEVs to matter in the $(\bm{4}, \bm{1},
\overline{\bm{2}})$ and  $(\bm{1}, \bm{3}, \overline{\bm{2}})$ representations
(as long as there is enough matter in these representations). Depending on
which bifundamental we give a VEV to first, we would have an intermediate
gauge algebra of either $\asu(3) \oplus \asu(3) \oplus \au(1)$ or $\asu(4)
\oplus \asu(2) \oplus \au(1)$. These intermediate stages should be visible
when unHiggsing the Weierstrass model. In fact, setting $s_2$ to zero, as done
above, should be viewed as a combination of two tunings. If we first let
$\tuning{s_2}{d_0 s_2^\prime}$, the gauge algebra enhances to $\asu(3) \oplus
\asu(3) \oplus \au(1)$. Subsequently setting $\tuning{s_2^\prime}{0}$ gives us
the full $\SMuh$ gauge group. Alternatively, we could first enhance the gauge
algebra to $\asu(4) \oplus \asu(2) \oplus \au(1)$ by letting $\tuning{s_2}{b_1
s_2^{\prime \prime}}$ and then set $\tuning{s_2^{\prime \prime}}{0}$. Either
way, the enhancement to $\SMuh$ can generally be described as a two step
process, as expected from field theory considerations.\footnote{Note that in
some cases, such as in 6D models where there is only a single hypermultiplet
in each of the three bifundamental representations, the Higgsing involves all
three bifundamentals and occurs in a single step.}

\section{Class (B) models}
\label{sec:class-B}

\subsection{Specialization to Class (B)}
\label{sec:spec-to-B}

The Weierstrass model described by \cref{eq:su321weierstrass} can also realize
the Class (B) models. 
Recall that $Y = [s_2]$ is trivial in these models (i.e., is in the zero class
in cohomology), while $\beta = [s_1]$ is allowed to be ineffective. Since
$s_1$ appears in the Weierstrass model, it naively seems that the construction
is invalid when $[s_1]$ is ineffective. But there is a way of salvaging the
model in these situations: if $[s_1]$ is ineffective, one can simply set $s_1$
to 0 in the Weierstrass model. Of course, setting a parameter to 0 can
introduce problems such as codimension-two $(4, 6)$ singularities, which are
generally associated with superconformal sectors in the theory  (see, e.g.,
\cite{HeckmanRudeliusSCFT} for a review), or an exactly vanishing
discriminant. For this situation, setting $s_1$ to zero does not introduce
these more serious issues, but the discriminant does become proportional to
$s_2^2$. This would suggest that we have an extra, undesired gauge group
unless $[s_2] = Y$ is trivial. Indeed, the analysis of \cite{TaylorTurner321}
states that $Y$ must be trivial for the Class (B) models. The Weierstrass
model directly reflects this fact and provides an explanation for the trivial
$Y$: unless $Y$ is trivial, the gauge group of the model is enhanced to
something larger than the standard model gauge group.

There is another way of seeing that $\beta = [s_1]$ can be ineffective as long
as $Y = [s_2]$ is trivial. When $Y$ is trivial, $s_2$ is essentially a nonzero
constant, so we can freely divide by $s_2$ without issue. With this in mind,
we can redefine the parameters $s_5$, $s_6$, and $s_8$ as
\begin{equation}
\label{eq:redefremoves1}
s_5 = s_5^\prime + \frac{s_1}{s_2^2} \left(s_2 s_6^\prime + b_1 d_0 s_1\right)\,, \quad s_6 = s_6^\prime + \frac{2}{s_2} b_1 d_0 s_1\,, \quad s_8 = s_8^\prime + \frac{1}{s_2} d_1 b_1 s_1\,.
\end{equation}
Note that, as long as $[s_2] = Y$ is trivial, these are simple redefinitions
involving shifts in the parameters $s_5$, $s_6$, $s_8$. But these
redefinitions remove all the terms in the Weierstrass model containing $s_1$.
In essence, the $s_1$ terms have been absorbed into the other parameters. And
since $s_1$ no longer appears in the Weierstrass model, clearly $s_1$ can be
ineffective.

However, it is known that the Class (B) spectra can also be realized by the
$F_{11}$ model described in \cite{KleversEtAlToric}. This would suggest that
the Weierstrass model above should match the $F_{11}$ model, at least when $Y$
is trivial. Indeed, we can redefine the parameters above in terms of the
parameters in the $F_{11}$ model:\footnote{We use the symbol $S_i$ to refer to
the parameter $s_i$ in the $F_{11}$ model.}
\begin{equation}
\begin{aligned}
d_2 &= \frac{1}{s_2^2} \left(S_1 - S_{11} S_2 + S_{11}^2 S_{3}\right)\,, & d_1 &= \frac{1}{s_2} \left(S_2 - S_3 S_{11}\right)\,, & d_0 &= S_3 \\
s_8^\prime &= \frac{S_5}{s_2}\,, & s_6^\prime &= S_6\,, & b_1 &= S_9\,.
\end{aligned}
\end{equation}
Here, $S_{11}$ is a new parameter not in the original $F_{11}$ model, defined
as
\begin{equation}
S_{11} = s_5^\prime\,.
\end{equation}
After these redefinitions, the $f$ for the Weierstrass model defined here
(when $Y$ is trivial) is exactly the same as the $f$ for the $F_{11}$ model.
The $g$'s, meanwhile, agree up to a term proportional to the new parameter
$S_{11}$:
\begin{equation}
g = g_{F_{11}} - S_3^2 S_9^3 S_{11} \left(S_1 - S_{11} S_2 + S_{11}^2 S_3\right)\,.
\end{equation}
This would suggest that the Weierstrass model presented here gives a slight
generalization of the $F_{11}$ model, with exact agreement when
$\tuning{S_{11}}{0}$.

The presence of this extra term should not be too surprising. If one were to
tune an $\SU(3) \times \SU(2)$ gauge group, a term in $g$ proportional to
$b_1^3 d_0^2$ (or $S_9^3 S_3^2$) would only contribute terms in $\Delta$ at
order $b_1^3 d_0^2$ or higher. Therefore, this extra term would not affect the
$\SU(3) \times \SU(2)$ tuning. Of course, it would have to take some special
form to allow for the extra section generating the $\U(1)$, and we indeed see
additional structure in the extra term above. Nevertheless, it is natural to
include this extra term in $g$, even though it does not appear in the $F_{11}$
model.

\subsection{Pati--Salam enhancement}
\label{sec:221-enhance}

Some of the Class (B) spectra should admit unHiggsings to Pati--Salam models
\cite{TaylorTurner321}, and as first found in \cite{KleversEtAlToric}, setting
the parameter $S_5$ in the $F_{11}$ model to 0 leads to an F-theoretic
Pati--Salam model. These two results suggest that our Weierstrass model should
similarly admit an enhancement to a Pati--Salam model. Indeed, the tuning
\begin{equation}
\tuning{s_8^\prime}{\frac{1}{s_2} s_5^\prime s_6^\prime}\,,
\end{equation}
makes the discriminant proportional to
\begin{equation}
b_1^4 d_0^2 \left(d_2 s_2^2 - d_1 s_5^\prime s_2 + d_0 {s_5^\prime}^2\right)^2\,.
\end{equation}
The corresponding gauge algebra  of the enhanced theory is then
$\asu(4) \oplus \asu(2) \oplus \asu(2)$. The $\asu(4)$ is tuned on a divisor of
class $b_3$, and the two $\asu(2)$s are tuned on divisors of classes $b_2$ and
$-4 a - 2 b_3 - b_2$. (Recall that $Y = [s_2]$ is trivial for the Class (B)
models we are currently considering.) This exactly matches the expectations
from \cite{TaylorTurner321}. Moreover, the $\U(1)$ generating section
\labelcref{eq:su321section} becomes
\begin{equation}
[\hat{x} : \hat{y} : \hat{z}] = \left[\frac{1}{12} \left[{s_6^\prime}^2 - 4 b_1 \left(d_1 s_2 - 2 d_0 s_5^\prime\right)\right] : 0 : 1\right]\,.
\end{equation}
Since the $\hat{y}$ component vanishes, the $\U(1)$ generating section has now
become a torsional section of order 2.\footnote{This phenomenon is similar to
those observed in \cite{BaumeEtAlTorsion}.} In fact, this new torsional section is essentially the same as that identified in  \cite{KleversEtAlToric}, with the only differences coming from terms proportional to $S_{11}$. If one performs an analysis similar to that in Appendix B of  \cite{KleversEtAlToric}, one
can conclude that the gauge group is $\PS$, exactly as expected.

\subsection{Enhancement to $\SU(5)$}
\label{sec:5-enhance}

The field theory analysis in \cite{TaylorTurner321} identified a second type of
enhancement that should be possible for some of the Class (B) models: when $Y = 0$ and $b_3 = b_2 =
-\canonclass$, the $\SM$ gauge group has a spectrum suggesting that the gauge group can be enhanced to an $\SU(5)$ group tuned
on a divisor of class $-\canonclass$. This unHiggsing is essentially the inverse process of
the Higgsing in the Georgi--Glashow GUT model \cite{GeorgiGlashow}, in which matter in the adjoint ($\bm{24}$) representation
of $\SU(5)$ obtains a VEV. Even though the Class (B) models can be realized by
the previous F-theory constructions in \cite{KleversEtAlToric}, no previous
work has, to the best of our knowledge, explicitly demonstrated that the
unHiggsing to $\SU(5)$ can be seen in a Weierstrass model. As we show below, the
Weierstrass construction in \cref{eq:su321weierstrass} explicitly admits an enhancement to
$\SU(5)$, although the exact tunings required are somewhat complicated and cannot be seen in the toric picture.

When $Y = 0$ and $b_3 = b_2 = -\canonclass$, the parameter $s_1$ is
ineffective, and one can remove $s_1$ from the Weierstrass model through the
redefinitions in \cref{eq:redefremoves1}. To simplify the discussion,
we also set the parameter $s_2$, which has a trivial divisor class, to
1.\footnote{Alternatively, one can remove $s_2$ from the Weierstrass model by letting
$d_1 \to s_2^{-1} d_1$, $s_8^\prime \to s_2^{-1} s_8^\prime$, and $d_2 \to
s_2^{-2} d_2$. We are allowed to rescale parameters by inverse powers of $s_2$
because $[s_2]$ is trivial.} With these simplifications, the Weierstrass model
is given by
\begin{equation}
\begin{aligned}
f &= -\frac{1}{48} \left[{s_6^\prime}^2 - 4 b_1 \left(d_0 s_5^\prime + d_1\right)\right]^2 + \frac{1}{2} b_1 d_0 \left[2 b_1 \left(d_1 s_5^\prime + d_2\right) - s_6^\prime s_8^\prime\right]\,, \\
g &= \frac{1}{864} \left[{s_6^\prime}^2 - 4 b_1 \left(d_0 s_5^\prime + d_1\right)\right]^3 + \frac{1}{4} b_1^2 d_0^2 \left({s_8^\prime}^2 - 4 b_1 d_2 s_5^\prime\right) \\
&\qquad - \frac{1}{24} b_1 d_0 \left[{s_6^\prime}^2 - 4 b_1 \left(d_0 s_5^\prime + d_1\right)\right] \left[2 b_1 \left(d_1 s_5^\prime + d_2\right) - s_6^\prime s_8^\prime\right]\,,
\end{aligned}
\end{equation}
while the generating section for the $\U(1)$ gauge factor is
\begin{equation}
[\secx : \secy : \secz] = \left[\frac{1}{12} \left[{s_6^\prime}^2 - 4 b_1 \left(d_1 -2 d_0 s_5^\prime\right)\right] : \frac{1}{2} b_1 d_0 \left(s_5^\prime s_6^\prime - s_8^\prime\right) : 1\right]\,.
\end{equation}
The parameters $s_6^\prime$, $s_8^\prime$, $b_1$, $d_0$, $d_1$, and $d_2$
are all sections of $\cO(-\canonclass)$,
while the parameter $s_5^\prime$ has a trivial divisor class.

One might expect that the gauge group enhances to $\SU(5)$ when the $\SU(2)$
locus coincides with the $\SU(3)$ locus. Operationally, one would tune $d_0$ to
be $b_1$, making the discriminant proportional to $b_1^5$. If this unHiggsing
truly is the inverse process of the adjoint Higgsing, we should see the $\U(1)$
gauge factor ``merge'' with the other nonabelian factors. At a practical level,
we would expect the $\secz$ component of the generating section to vanish such
that the generating section coalesces with the zero section.  But if we naively
set $\tuning{d_0}{b_1}$ in the above expressions, the generating section remains
distinct from the zero section. Tuning $\tuning{d_0}{b_1}$ therefore represents a
different enhancement. Since the generating section remains after setting $\tuning{d_0
}{b_1}$, the enhanced gauge algebra is $\asu(5) \oplus \au(1)$, not
$\asu(5)$.\footnote{The section cannot be a $\Z_5$ torsional section
  in an F-theory model without codimension-two (4, 6) loci, as can be
  seen from the general form of a Weierstrass model with such a
  section \cite{AspinwallMorrisonNonsimply,
    MorrisonTaylorCompleteness}.} This enhancement cannot be the unHiggsing that we want, as we should not
see any extra $\U(1)$ factors after the tuning.

The correct $\SU(5)$ enhancement procedure still involves tuning the $\SU(3)$
and $\SU(2)$ loci to coincide, but the exact tuning is more subtle. It is
easiest to state the procedure before describing the underlying logic. First,
we let
\begin{equation}
\label{eq:su5enhsu2div}
d_0 = b_1 - \epsilon \dzerotilde\,,
\end{equation}
where $[\epsilon] = 0$ and $[\dzerotilde] = -\canonclass$. This
redefinition does not, by itself, restrict the structure of $d_0$. Since
$[\epsilon] = 0$, we have simply performed a shift and rescaling of $d_0$. And
we can always undo this by letting $\dzerotilde = \epsilon^{-1}(-d_0 + b_1)$.
Next, we redefine the other parameters in terms of $\epsilon$:
\begin{gather}
s_5^\prime = \frac{1}{\epsilon^2} + \tilde{s}_5\,, \label{eq:su5enhredef1} \\
s_6^\prime = -\dzerotilde + \epsilon \tilde{s}_6\,, \\
s_8^\prime = -\frac{2}{\epsilon^3} b_1 - \frac{2}{\epsilon} b_1 \tilde{s}_5 + \tilde{d}_1 + \epsilon \tilde{s}_8\,, \\
d_1 = -\frac{1}{\epsilon^2} \left(b_1 - \epsilon \dzerotilde\right) - \tilde{s}_6 + \epsilon \tilde{d}_1\,, \\
d_2 = \frac{1}{\epsilon^4} b_1 + \frac{1}{\epsilon^2} \tilde{s}_5 b_1 - \frac{1}{\epsilon} \tilde{d}_1 - \tilde{s}_8 + \epsilon^2 \tilde{d}_2\,. \label{eq:su5enhredef5}
\end{gather}
The Weierstrass model is now described by
\begin{equation}
\begin{aligned}
f &= -\frac{1}{48} \left\{(\dzerotilde - \epsilon \tilde{s}_6)^2 - 4 b_1 \left[\tilde{s}_5 (b_1 - \epsilon \dzerotilde) - (\tilde{s}_6 - \epsilon \tilde{d}_1)\right]\right\}^2 \\
&\qquad - \frac{1}{2} b_1 (b_1 - \epsilon \dzerotilde) \left[2 b_1 \left(\tilde{s}_8 - \epsilon \tilde{d}_1 \tilde{s}_5 - \epsilon^2 \tilde{d}_2\right) - (\dzerotilde - \epsilon \tilde{s}_6) (\tilde{d}_1 + \epsilon \tilde{s}_8)\right]
\end{aligned}
\end{equation}
and
\begin{equation}
\begin{aligned}
g &= \frac{1}{864} \left\{(\dzerotilde - \epsilon \tilde{s}_6)^2 - 4 b_1 \left[\tilde{s}_5 (b_1 - \epsilon \dzerotilde) -(\tilde{s}_6 - \epsilon \tilde{d}_1)\right]\right\}^3 \\
&\qquad + \frac{1}{4} b_1^2 (b_1 - \epsilon \dzerotilde)^2 \left[(\tilde{d}_1 + \epsilon \tilde{s}_8)^2 - 4 b_1 \left(\tilde{d}_2 + \epsilon^2 \tilde{d}_2 \tilde{s}_5\right)\right] \\
&\qquad + \frac{1}{24} b_1 (b_1 - \epsilon \dzerotilde) \left\{(\dzerotilde - \epsilon \tilde{s}_6)^2 - 4 b_1 \left[\tilde{s}_5 (b_1 - \epsilon \dzerotilde) - (\tilde{s}_6 - \epsilon \tilde{d}_1)\right]\right\}  \\
&\qquad\qquad\qquad\qquad\qquad\quad \times \left[2 b_1 \left(\tilde{s}_8 - \epsilon \tilde{d}_1 \tilde{s}_5 - \epsilon^2 \tilde{d}_2\right) - (\dzerotilde - \epsilon \tilde{s}_6) (\tilde{d}_1 + \epsilon \tilde{s}_8)\right]\,.
\end{aligned}
\end{equation}
Miraculously, $f$ and $g$ do not contain any terms proportional to inverse
powers of $\epsilon$, even though the redefinitions above include inverse
powers of $\epsilon$. The generating section components, on the other hand, do
have inverse powers of $\epsilon$ after the redefinitions, but we can remove
these inverse powers by rescaling the section components. In the end, the
generating section is given by $[\secx : \secy : \secz]$,
with
\begin{equation}
\begin{aligned}
\secx &= \frac{1}{12} \epsilon^2 \left(\dzerotilde - \epsilon \tilde{s}_6\right)^2 + b_1^2 + \frac{2}{3} \epsilon^2 \tilde{s}_5 b_1^2 - \frac{1}{3} \epsilon b_1 \left[3 \dzerotilde - \epsilon \tilde{s}_6 + \epsilon^2 \left(\tilde{d}_1 + 2 \dzerotilde \tilde{s}_5\right)\right]\,, \\
\secy &= -\frac{1}{2} b_1 \left(b_1 - \epsilon \dzerotilde\right) \left[\epsilon^3 \tilde{d}_1 + \epsilon^4 \tilde{s}_8 - \left(1 + \epsilon^2 \tilde{s}_5\right) \left(2 b_1 - \epsilon \dzerotilde + \epsilon^2 \tilde{s}_6\right)\right]\,, \\
\secz &= \epsilon\,.
\end{aligned}
\end{equation}

Since $f$ and $g$ contain no inverse powers of $\epsilon$, we can now safely
send $\tuning{\epsilon}{0}$. We find that $f$ and $g$ do not vanish in this
limit, but the discriminant becomes proportional to $b_1^5$. One can also
verify that $f$ and $g$ satisfy the split condition, indicating that we have a
split $I_5$ singularity along $\locus{b_1 = 0}$. Thus, we have an $\SU(5)$ tuned on a
divisor of class $-\canonclass$, as expected. However, the $\secz$ component
of the section now vanishes in the $\tuning{\epsilon}{0}$ limit, and the
generating section coalesces with the zero section. There are no extra $\U(1)$
factors, and the enhanced gauge group is $\SU(5)$. The $\tuning{\epsilon}{0}$
tuning is therefore the desired unHiggsing process.

The tuning procedure described above admittedly seems ad-hoc, but there is an
underlying logic behind at least some of the steps. In particular, the need
for the new parameter $\epsilon$ can be understood from the $\SU(5) \to \SM$
branching rules:\footnote{Note that the representations listed in the
branching rules below may differ by conjugation from those listed in
\cref{tab:codimtwoloci}.}
\begin{equation}
\begin{aligned}
\bm{24} &\to (\bm{8}, \bm{1})_0 + (\bm{1}, \bm{3})_0 + (\bm{3}, \bm{2})_{-5 / 6} + (\overline{\bm{3}}, \bm{2})_{5 / 6} + (\bm{1}, \bm{1})_0\,, \\
\bm{10} &\to (\bm{3}, \bm{2})_{1 / 6} + (\overline{\bm{3}}, \bm{1})_{-2 / 3} + (\bm{1}, \bm{1})_1\,, \\
\bm{5} &\to (\bm{3}, \bm{1})_{-1 / 3} + (\bm{1}, \bm{2})_{1 / 2}\,.
\end{aligned}
\end{equation}
The branching rules involve two types of bifundamentals: those with charge $q
= \pm\frac{5}{6}$, and those with charge $q = \frac{1}{6}$. Of course, the
matter spectrum for the situation at hand would not actually contain any
$(\bm{3}, \bm{2})_{-5 / 6}$ matter, since, in the actual Higgsing process, all the
$(\bm{3}, \bm{2})_{-5 / 6}$ matter is eaten by the broken gauge
bosons.\footnote{This occurs because the $\SU(5)$ is tuned on a divisor of
class $-\canonclass$, implying that, at least in a 6D model, the $\SU(5)$
matter spectrum contains only one hypermultiplet of $\bm{24}$ matter. If we
considered a model with more than one $\SU(5)$ adjoint hypermultiplet, the
resulting spectrum would contain $(\bm{3}, \bm{2})_{-5 / 6}$ hypermultiplets.}
Nevertheless, if we want to see the $\SU(5)$ unHiggsing process explicitly,
the Weierstrass model should contain a ``would-be'' $(\bm{3}, \bm{2})_{-5 /
6}$ locus. Since bifundamentals come from intersections between gauge
divisors, a construction that demonstrates an $\SU(5)$ unHiggsing should
present two ways of having the $\SU(3)$ and $\SU(2)$ loci coincide. This motivates
the specific form of $d_0$ in \cref{eq:su5enhsu2div}, as we can send
$\tuning{d_0}{b_1}$ by letting $\tuning{\dzerotilde}{0}$ or by letting
$\tuning{\epsilon}{0}$. Here, $\locus{b_1 = \epsilon = 0}$ is the would-be $(\bm{3},
\bm{2})_{-5 / 6}$ locus, since $[\epsilon] = 0$. We also need to ensure that
as $\tuning{\epsilon}{0}$, the generating section coalesces with the zero
section. This implies that we need to redefine terms so that the $\secz$ component
of the section becomes proportional to $\epsilon$. In the end, we need $\SU(3)$ and $\SU(2)$ gauge factors tuned on divisors
taking the forms described above, and we need a specific form of the $\secz$
component of the section. One can use these requirements to find the necessary
redefinitions of the parameters, leading to
\crefrange{eq:su5enhredef1}{eq:su5enhredef5}.

Finally, we note that even though the enhancement was described for the
Weierstrass construction presented here, a similar $\SU(5)$ unHiggsing should
be possible in the $F_{11}$ construction of \cite{KleversEtAlToric}. As
mentioned previously, there is a map between these two constructions when $Y =
0$. It is therefore possible to use this dictionary to find the necessary
tunings in the $F_{11}$ construction, although we do not go through the
details of this analysis here.

\section{Codimension-two singularities and matter}
\label{sec:matter}

\begin{table}
\centering

\[\begin{array}{ccc}\toprule
\text{Locus}                                          & \text{Multiplicity}                                   & \text{Supported Matter} \\ \midrule
\locus{b_1 = d_0 = 0}                                 & b_3 \cdot b_2                                         & (\bm{3}, \bm{2})_{\mathrlap{1 / 6}} \\
\locus{b_1 = s_2 = 0}                                 & b_3 \cdot Y                                           & (\bm{3}, \bm{1})_{\mathrlap{-4 / 3}} \\
\locus{b_1 = d_2 s_6^2 - d_1 s_6 s_8 + d_0 s_8^2 = 0} & b_3 \cdot X                                           & (\bm{3}, \bm{1})_{\mathrlap{-1 / 3}} \\
\locus{b_1 = s_8 s_2^2 - s_5 s_2 s_6 + s_1 s_6^2 = 0} & b_3 \cdot \left(\beta - 2\canonclass\right)           & (\bm{3}, \bm{1})_{\mathrlap{2 / 3}} \\
\locus{d_0 = \sutwofundlocus = 0}                     & b_2 \cdot \left(X + \beta - \canonclass\right)        & (\bm{1}, \bm{2})_{\mathrlap{1 / 2}} \\
\locus{d_0 = s_2 = 0}                                 & b_2 \cdot Y                                           & (\bm{1}, \bm{2})_{\mathrlap{3 / 2}} \\
\chargeonelocus                                       & (b_3 + b_2 + 2 \beta) \cdot X - \canonclass \cdot b_2 & (\bm{1}, \bm{1})_{\mathrlap{1}} \\
\locus{s_2 = s_1 = 0}                                 & \beta \cdot Y                                         & (\bm{1}, \bm{1})_{\mathrlap{2}} \\\bottomrule
\end{array}\]

\caption{Codimension-two loci of the $\SM$ model and the associated charged
matter. The multiplicities are for a 6D model constructed using the $\SM$
Weierstrass model \labelcref{eq:su321weierstrass}. Note that $\sutwofundlocus$
is defined in \cref{eq:su2fundlocusdef}, while $\chargeonelocus$ is defined in
\cref{eq:chargeonelocusdef}.}
\label{tab:codimtwoloci}
\end{table}

There are two sources of charged matter in the F-theory model described by
\cref{eq:su321weierstrass}. First, the model can have matter in the adjoint
representation of $\SU(3)$ or $\SU(2)$ coming from strings that propagate
freely along the gauge divisors. In 6D F-theory models, where the gauge
divisors are complex curves in the base, adjoint matter occurs when either the
$\SU(3)$ divisor $\locus{b_1 = 0}$ or the $\SU(2)$ divisor $\locus{d_0 = 0}$
has a genus $g$ greater than zero.\footnote{We are assuming here that the
divisors are smooth. When the divisors have singularities, there may be
adjoint matter localized at the singular loci, or the singularities may
support matter in more exotic representations \cite{SadovGreenSchwarz,MorrisonTaylorMaS,KleversEtAlExotic}.} The number of adjoint
hypermultiplets is given by the geometric genus of the corresponding gauge
divisors. Specifically, there are $1 + b_3 \cdot (b_3 + \canonclass) / 2$
hypermultiplets of $(\bm{8}, \bm{1})_0$ matter and $1 + b_2 \cdot (b_2 +
\canonclass) / 2$ hypermultiplets of $(\bm{1}, \bm{3})_0$ matter. These
adjoint hypermultiplets are uncharged under the $\U(1)$ and are charged only
under one of the nonabelian gauge group factors.

Matter can also be supported along codimension-two loci in the base where the
elliptic curve singularity type enhances. The codimension-two loci supporting
matter in the $\SM$ model are summarized in \cref{tab:codimtwoloci}, but let
us describe the process of finding these loci and the associated matter
representations in more detail.

\subsection{Determining the matter loci}
\label{sec:loci}

First, let us focus on the codimension-two loci along the $\SU(3)$ divisor
$\locus{b_1 = 0}$. Any matter supported at such loci should be charged under
the $\SU(3)$ gauge group. The discriminant takes the form
\begin{equation}
\Delta = b_1^3 d_0^2 \left[\frac{1}{16} s_2 s_6^3 \left(s_8 s_2^2 - s_5 s_2 s_6 + s_1 s_6^2\right) \left(d_2 s_6^2 - d_1 s_6 s_8 + d_0 s_8^2\right) + \cO(b_1)\right]\,.
\end{equation}
From this expression, we see five different codimension-two loci along
$\locus{b_1 = 0}$ where the singularity type enhances.  At $\locus{b_1 = d_0 =
0}$, the intersection locus of the $\SU(3)$ and $\SU(2)$ divisors, the
singularity type enhances from $I_3$ to $I_6$. As we will see shortly, this
locus supports bifundamental matter. At $\locus{b_1 = s_6 = 0}$, meanwhile,
the singularity type changes from $I_3$ to $IV$.  Such loci do
not contribute charged matter, since both describe $\SU(3)$ type
singularities, so we can ignore $\locus{b_1 =
s_6 = 0}$ for the purposes of the charged matter analysis. We are left with
three loci where the singularity type enhances from $I_3$ to $I_4$:
$\locus{b_1 = s_2 = 0}$, $\locus{b_1 = s_8 s_2^2 - s_5 s_2 s_6 + s_1 s_6^2 =
0}$, and $\locus{b_1 = d_2 s_6^2 - d_1 s_6 s_8 + d_0 s_8^2 = 0}$. These loci
support matter in the fundamental representation of $\SU(3)$ with different
$\U(1)$ charges.

Let us now turn to the codimension-two loci along $\locus{d_0 = 0}$, which should support matter charged under the $\SU(2)$ gauge factor. The discriminant can be written as
\begin{equation}
\Delta = b_1^3 d_0^2\left[-\frac{1}{16} s_2 \left(s_6^2 - 4 b_1 d_1 s_2\right)^2 \sutwofundlocus + \cO(d_0)\right]\,,
\end{equation}
where
\begin{equation}
\begin{aligned}
\label{eq:su2fundlocusdef}
\sutwofundlocus &= b_1^2 d_1^3 s_1^2 + b_1 \left[d_2^2 s_2^3 + d_1 d_2 \left(3 s_1 s_6 - 2 s_2 s_5\right) s_2 + d_1^2 \left(s_2 s_5^2 - s_1 s_6 s_5 - 2 s_1 s_2 s_8\right)\right] \\
&\qquad - \left(s_8 s_2^2 - s_5 s_6 s_2 + s_1 s_6^2\right) \left(d_2 s_6 - d_1 s_8\right)\,.
\end{aligned}
\end{equation}
The locus $\locus{b_1 = d_0}$, where the singularity type enhances from $I_2$
to $I_6$, was mentioned previously. At the locus $\locus{d_0 = s_6^2 - 4 b_1
d_1 s_2 = 0}$, the singularity type enhances to type $III$. This locus, just
like the $\locus{b_1 = s_6 = 0}$ locus discussed before, does not contribute
charged matter since both singularity types are associated with $\SU(2)$
groups, so we will not discuss it further. We are therefore left with two loci
where the singularity type enhances from $I_2$ to $I_3$: $\locus{d_0 = s_2 =
0}$ and $\locus{d_0 = \sutwofundlocus = 0}$. These loci support matter in the
fundamental representation of $\SU(2)$.

Finally, we need to determine the codimension-two matter loci not along
$\locus{b_1 = 0}$ or $\locus{d_0 = 0}$. Such loci, which support matter
charged only under the $\U(1)$ gauge group, are contained in
\begin{equation}
\secy = 3 \secx^2 + f \secz^4 = 0\,,
\end{equation}
where $[\secx : \secy : \secz]$ are the section components of the generating
section. Of course, we want to focus on the subloci not involving $\locus{b_1
= 0}$ or $\locus{d_0 = 0}$. After some analysis, one finds that the
appropriate sublocus containing the desired matter loci is
\begin{equation}
V = \left\{\begin{array}{l}\phantom{=}-b_1 d_1 s_1 s_2^3 + 2 b_1 d_0 s_1 s_5 s_2^2 - 3 b_1 d_0 s_1^2 s_6 s_2 + 2 b_1^2 d_0^2 s_1^3 + s_8 s_2^4 - s_5 s_6 s_2^3 + s_1 s_6^2 s_2^2 \\
= - b_1 d_0 s_1^2 s_2 \left(d_1 s_2^2 - 2 d_0 s_5 s_2 + 2 d_0 s_1 s_6\right) + b_1^2 d_0^3 s_1^4 + d_2 s_2^6 \\
\qquad\qquad\qquad\qquad - s_2^2 \left(s_2 s_5 - s_1 s_6\right) \left(d_1 s_2^2 - d_0 s_5 s_2 + d_0 s_1 s_6\right) = 0\end{array}\right\}\,.
\end{equation}

There are two important subloci of $V$. The first is the sublocus $\locus{s_1
= s_2 = 0}$, along which the $[\secx : \secy : \secz]$ components vanish. We
will later show that this sublocus supports matter in the  $(\bm{1},
\bm{1})_2$ representation. The other important sublocus is essentially $V$
with all of the previous matter loci removed. In particular, $V$ contains the
loci $\locus{s_1 = s_2 = 0}$, $\locus{b_1 = s_2 = 0}$, and $\locus{d_0 = s_2 =
0}$, so the locus we are interested in is
\begin{equation}
\label{eq:chargeonelocusdef}
\chargeonelocus = V \setminus \left(\locus{s_1 = s_2 = 0} \cup \locus{b_1 = s_2 = 0} \cup \locus{d_0 = s_2 = 0}\right)\,.
\end{equation}
As might be anticipated by its name, $\chargeonelocus$ supports matter in the
$(\bm{1}, \bm{1})_1$ representation. Determining the multiplicity for
$\chargeonelocus$ is somewhat complicated. Based on $V$, one might naively
expect that the multiplicity for $V$ is $\left(2 b_3 + 2 b_2 + 3 \beta\right)
\cdot \left(2 b_3 + 3 b_2 + 4 \beta\right)$. But we must account for the
contributions from the undesired loci $\locus{s_1 = s_2 = 0}$, $\locus{b_1 =
s_2 = 0}$, and $\locus{d_0 = s_2 = 0}$. As part of this analysis, we need to
determine how many copies of the undesired loci are contained within $V$. If
we write $V$ as $\locus{v_a = v_b = 0}$, this information can be determined
from the resultant of $v_a$ and $v_b$ with respect to $s_2$
\cite{CveticKleversPiraguaMultU1}. One finds that
\begin{equation}
\Res_{s_2}(v_a, v_b) \propto b_1^8 d_0^9 s_1^{16}\,,
\end{equation}
indicating that $V$ contains eight copies of $\locus{b_1 = s_2 = 0}$, nine
copies of $\locus{d_0 = s_2 = 0}$, and sixteen copies of $\locus{s_1 = s_2 =
0}$. Therefore, the multiplicity for $\chargeonelocus$ is
\begin{equation}
\begin{aligned}
&\left(2 b_3 + 2 b_2 + 3 \beta\right) \cdot \left(2 b_3 + 3 b_2 + 4 \beta\right) - (8 b_3 + 9 b_2 + 16 \beta) \cdot Y \\
&\qquad\qquad\qquad = (b_3 + b_2 + 2 \beta) \cdot \left(-8 \canonclass - 4 b_3 - 3 b_2 - 2 \beta\right) - \canonclass \cdot b_2\,.
\end{aligned}
\end{equation}

\subsection{Determining the matter representations}

Now that we have identified the important codimension-two loci, we can
investigate the types of charged matter supported at these loci. While a
proper analysis would require resolving singularities in the Calabi--Yau
manifold, we can also determine the supported representations more quickly
using heuristic arguments. In particular, the Katz--Vafa \cite{KatzVafa} method tells us about the nonabelian representations of the
supported matter. And we can use our knowledge of the unHiggsing patterns and
the relations to previously found models to determine the $\U(1)$ charges of
the matter.

First, recall that we can exactly recover the $\SM$ model of
\cite{KleversEtAlToric} by setting $s_1$ and $s_5^\prime$ to zero and $s_2$ to
a constant. Therefore, the matter loci in these two models should match after
the appropriate variable specializations have been made. In this situation,
the only relevant codimension-two loci are $\locus{b_1 = d_0 = 0}$,
$\locus{b_1 = s_8 s_2^2 - s_5 s_2 s_6 + s_1 s_6^2 = 0}$, $\locus{b_1 = d_2
s_6^2 - d_1 s_6 s_8 + d_0 s_8^2 = 0}$, $\locus{d_0 = \sutwofundlocus = 0}$,
and $\chargeonelocus$. (The other loci disappear once $s_2$ is set to a
nonzero constant.) We can match each of these loci to loci in
\cite{KleversEtAlToric} and determine the matter representations:
\begin{itemize}
\item
The locus $\locus{b_1 = d_0 = 0}$ is the intersection between the two gauge
divisors and should therefore support bifundamental matter. In
\cite{KleversEtAlToric}, the corresponding locus supports $(\bm{3}, \bm{2})_{1
/ 6}$ matter,\footnote{In \cite{KleversEtAlToric}, the charge is listed as $-1
/ 6$, and the signs of the charges listed are in general the negatives of
those listed in \cite{TaylorTurner321}. Since the overall sign of the charges
can be flipped freely, we flip the signs of the charges to match
\cite{TaylorTurner321}.} so $\locus{b_1 = d_0 = 0}$ should also support
$(\bm{3}, \bm{2})_{1 / 6}$ matter.
\item
At the locus $\locus{b_1 = s_8 s_2^2 - s_5 s_2 s_6 + s_1 s_6^2 = 0}$, the
singularity type enhances from $I_3$ to $I_4$. This locus therefore supports
matter in the fundamental representation of $\SU(3)$. After setting $s_1$ to
zero and $s_2$ to a constant, this locus corresponds to the locus $\locus{S_9
= S_5 = 0}$ in \cite{KleversEtAlToric}, which supports $(\bm{3}, \bm{1})_{2 /
3}$ matter. The locus $\locus{b_1 = s_8 s_2^2 - s_5 s_2 s_6 + s_1 s_6^2 = 0}$
therefore also supports $(\bm{3}, \bm{1})_{2 / 3}$ matter.
\item
At the locus $\locus{b_1 = d_2 s_6^2 - d_1 s_6 s_8 + d_0 s_8^2 = 0}$ , the
singularity type enhances from $I_3$ to $I_4$. This locus therefore supports
matter in the fundamental representation of $\SU(3)$. After setting $s_1$ and
$s_5^\prime$ to zero and $s_2$ to a constant, this locus corresponds to the
locus $\locus{S_9 = S_3 S_5^2 - S_2 S_5 S_6 + S_1 S_6^2 = 0}$ in
\cite{KleversEtAlToric}, which supports $(\bm{3}, \bm{1})_{-1 / 3}$ matter.
The locus $\locus{b_1 = d_2 s_6^2 - d_1 s_6 s_8 + d_0 s_8^2 = 0}$ therefore
supports  $(\bm{3}, \bm{1})_{-1 / 3}$ matter.
\item
At the locus $\locus{d_0 = \sutwofundlocus = 0}$ , the singularity type
enhances from $I_2$ to $I_3$, indicating that this locus supports matter in
the fundamental representation of $\SU(2)$. After setting $s_1$ and
$s_5^\prime$ to zero and $s_2$ to a constant, this locus corresponds to the
locus $\locus{S_3 = S_2 S_5^2 - S_6 S_5 S_1 + S_9 S_1^2 = 0}$ in
\cite{KleversEtAlToric}, which supports $(\bm{1}, \bm{2})_{1 / 2}$ matter. The
locus $\locus{d_0 = \sutwofundlocus = 0}$ therefore supports $(\bm{1}, \bm{2})_{1 /
2}$ matter.
\item
At the locus $\chargeonelocus$, the singularity type enhances to $I_2$. This
locus therefore supports matter charged under only the $\U(1)$ gauge factor.
After setting $s_1$ and $s_5^\prime$ to zero and $s_2$ to a constant, this
locus corresponds to the locus $\locus{S_1 = S_5 = 0}$  in
\cite{KleversEtAlToric}, which supports $(\bm{1}, \bm{1})_1$ matter. The locus
$\chargeonelocus$ therefore supports $(\bm{1}, \bm{1})_1$ matter.
\end{itemize}

There are three remaining codimension-two loci that do not have counterparts
in the model from \cite{KleversEtAlToric}: $\locus{s_1 = s_2 = 0}$,
$\locus{b_1 = s_2 = 0}$, and $\locus{d_0 = s_2 = 0}$. But there are
alternative ways of determining the matter representations of these loci
without performing a resolution. Because $\locus{s_1 = s_2 = 0}$ does not
involve either of the gauge divisors, the matter supported here can only be
charged under the $\U(1)$ gauge factor. In fact, the $[\secx : \secy : \secz]$
section components vanish to orders $(2, 3, 1)$ at $\locus{s_1 = s_2 = 0}$. In
models with just a $\U(1)$ gauge group, the section components vanish to these
orders at loci supporting charge $q = 2$ matter \cite{MorrisonParkU1,
Raghuram34}. One would therefore expect that $\locus{s_1 = s_2 = 0}$
supports $(\bm{1}, \bm{1})_2$ matter. Note that we have used the fact that the
Shioda map gives us a standard unit for the charge: matter with $\U(1)$ charge
$q = 1$ that is not charged under the non-abelian gauge factors occurs at
codimension two $I_2$ loci where the generating section intersects the extra
component transversely.

This leaves us with the loci $\locus{b_1 = s_2 = 0}$ and $\locus{d_0 = s_2 =
0}$. At $\locus{b_1 = s_2 = 0}$, the singularity type enhances from $I_3$ to
$I_4$, so the matter supported here must be in the fundamental representation
of $\SU(3)$. At $\locus{d_0 = s_2 = 0}$, meanwhile, the singularity type
enhances from $I_2$ to $I_3$, so the matter supported here should be in the
fundamental representation of $\SU(2)$. However, we still need to determine
the $\U(1)$ charges of these two types of matter. Fortunately, we know that
our model can be obtained by Higgsing a model with an $\SMuh$ gauge symmetry,
at least when $s_1$ is effective. Specifically, we give VEVs to bifundamental
matter in the $(\bm{4}, \bm{1}, \overline{\bm{2}})$ and $(\bm{1}, \bm{3},
\overline{\bm{2}})$ representations. If one works out how the $\SMuh$
representations branch to $\SM$ representations, one recovers all of the
representations mentioned so far, but one additionally finds the
representations $(\bm{3}, \bm{1})_{-4 / 3}$ and $(\bm{1}, \bm{2})_{3 / 2}$.
These must be the representations associated with the two remaining loci,
$\locus{b_1 = s_2 = 0}$ and $\locus{d_0 = s_2 = 0}$. Thus, $\locus{b_1 = s_2 =
0}$ should support $(\bm{3}, \bm{1})_{-4 / 3}$ matter, while $\locus{d_0 = s_2
= 0}$ should support $(\bm{1}, \bm{2})_{3 / 2}$.

This gives us the results summarized in \cref{tab:codimtwoloci}. The spectrum
satisfies the 6D anomaly cancellation conditions with the appropriate anomaly
coefficients, which gives us some confidence that we have obtained the correct
matter spectrum. Moreover, the spectrum agrees exactly with that described
\cite{TaylorTurner321}, further supporting the assertion that the Weierstrass
construction in \cref{eq:su321weierstrass} is the expected $\SM$ model.

As discussed to some extent already in \cite{TaylorTurner321}, the matter
spectrum distinguishes the Class (A) and (B) models.  The class (B) models,
with $Y = 0$ have precisely the matter representations expected for a
supersymmetric extension of the standard model, both in 6D and in 4D.  While
the Class (A) models are more general (parameterized by three divisor classes
rather than two), additional matter is potentially supported in these models.
In supersymmetric 6D models there are additional matter fields, specifically
the fields with charges $(\bm{1}, \bm{2})_{3 / 2}$, $(\bm{3}, \bm{1})_{-4 /
3}$, and $(\bm{1}, \bm{1})_2$.  In 4D, there is also the possibility of massless chiral matter in these representations. However, since chiral matter
is produced by fluxes, there are also consistent supersymmetric 4D F-theory
models with only the usual MSSM fields in the massless spectrum.
Understanding how much tuning is needed to avoid light exotics of these
representations, and how the spectrum is affected by supersymmetry breaking,
are interesting phenomenological questions left for further work.

\section{Full dimensionality of model}
\label{sec:dimension}

Given a parameterized Weierstrass model for F-theory constructions
with a given gauge group $G$ and associated generic matter,
of the type constructed in
\cref{sec:construction}, we would like to know if this is in fact the
most general such Weierstrass model, or if we are missing some
parameters that could be included in a more complete model.
There are several
ways of testing this, depending upon the situation.

In the simplest cases, with a single nonabelian gauge group factor
associated with fixed orders of vanishing of the Weierstrass
coefficients $f, g$ through the Kodaira classification (e.g.\ $E_6$),
it is straightforward to check directly that the Weierstrass model is
generic subject to those conditions.  This is slightly more subtle
when the gauge group requires an order of vanishing of the
discriminant $\Delta$ that is larger than that required by the orders
of vanishing of $f, g$, but a direct analysis in these cases is
relatively straightforward for a single gauge group factor.  For
example, in \cite{MorrisonTaylorMaS} the general form of Weierstrass
models for $\SU(N)$ to be tuned (over a smooth curve) is determined by
explicitly checking the Kodaira conditions for $f, g, \Delta$ in an
order-by-order expansion in the parameter $\sigma$ associated with the
$\SU(N)$ locus. Up to SU(5) this gives Weierstrass models associated
with the standard Tate tuning procedure
 (above SU(5) there
are multiple distinct branches of parameterized Weierstrass models,
including non-generic matter in the 3-index antisymmetric
representation of SU(6) through SU(8), see also
\cite{AndersonGrayRaghuramTaylorMiT, HuangTaylorLargeHodge}).

When the gauge group has abelian $\U(1)$
or multiple $\SU(N)$ factors, this question becomes more complicated.  In particular, we do
not know of a simple algebraic condition on the components of $f, g,
\Delta$ associated with the existence of a nontrivial section
associated with nonzero Mordell--Weil rank, as is needed for a $\U(1)$
factor.  As the simplest example of this, consider the Morrison--Park
model \cite{MorrisonParkU1} for a theory with U(1) gauge group and
generic matter charges $q = 1, 2$
\begin{equation}
\label{eq:morrison-park}
y^2 = x^3 + \left(c_1 c_3 - b^2 c_0 - \frac{1}{3} c_2^2\right) x + \left(c_0 c_3^2 - \frac{1}{3} c_1 c_2 c_3 + \frac{2}{27} c_2^3 - \frac{2}{3} b^2 c_0 c_2 + \frac{1}{4} b^2 c_1^2\right)\,.
\end{equation}
Here, $c_j$ is a section of a line bundle in the class $-2 \canonclass - (j - 2)
(\canonclass + L)$ where $\canonclass$ is the canonical class of the base and the line
bundle $L$ parameterizes the spectra ($b$ is a section of the line
bundle $- 2 \canonclass - L$).  In this case, the number of distinct complex
coefficients needed to choose the sections $c_i, b$ is in general
significantly larger than the number of moduli in the expected moduli
space; i.e., the parameterization is redundant.  In this kind of
situation it is a bit more subtle to check that the model indeed spans
a space of the proper dimensionality.  This question can be answered
most easily for such parameterized models in the context of
six-dimensional theories, where the dimension of the space of
Weierstrass models must match the number of uncharged hypermultiplets,
which is in turn fixed by anomaly cancellation.  For the Morrison--Park
model, one approach to counting the number of moduli was given in
\cite{WangU1s}.  Here we use a somewhat more direct method that works
for any such parameterized  Weierstrass model including those constructed in
\cref{sec:construction}.

The basic idea is fairly straightforward.  If we write the set of
complex coefficients of the component monomials in $f, g$ as $W_k \in \C$, and the set of
coefficients of the component monomials in the parameters $c_i, b$ of the Morrison--Park model
as $v_j \in \C$, then the dimension of the moduli space around a fixed
chosen background configuration is given by subtracting the dimension
of the set of automorphism symmetries from the rank (rk $J =$ dim Im
$J$) of the Jacobian matrix
\begin{equation}
 J_{k j} = \partial W_k / \partial v_j \,. \label{eq:jacobianmatrix}
\end{equation}
Since in the Morrison--Park model $f$ and $g$ are algebraic functions
of the $c_i, b$, each of the elements of the Jacobian matrix is simply
a polynomial in the $v_j$.  Because the dimensions of the matrix are
quite large, however, it is computationally difficult to check the
rank algebraically.  We proceed therefore numerically.  To avoid
precision issues, we simply choose  $\hat{v}_j$ to be random integers in
a given range $\hat{v}_j \in \{1, 2, \dots, N\}$, and then the rank of the
Jacobian matrix
\begin{equation}
{\rm rk}\ J|_{v = \hat{v}}
\end{equation}
 in the vicinity of the specific model with $v_j = \hat{v}_j$
is straightforward to compute using Mathematica or
another computational tool.  If the range $N$ of allowed integers is
sufficiently large, there will be many relatively prime factors in the distinct
$\hat{v}_j$ and the chances of a coincidental decrease in rank becomes very
small.  To check that this does not occur we have tested various
cases with the $\hat{v}_j$ varying in ranges from 1 to $N =$ 10 and
from 1 to $N =$ 1000
and in all cases we get the same answer, so empirically the range does
not need to be too large to get a correct measure of the rank.  As a specific
example, if we take the Morrison--Park model in the case where the base
is $\bP^2$, so $- \canonclass = 3 H$ where $H$ is the generating class (a line
in $\bP^2$), and we pick $L =2 H$, then there are 115 variables $v_j$,
and we expect the dimension of the moduli space to be 106 by anomaly
cancellation. (This is the case in which the dimension of the moduli
space is smallest).  Computing the rank using the above algorithm
gives 114. There is a redundancy in 8 degrees of freedom because of
possible reparameterization of the homogeneous coordinates on $\bP^2$
(up to an overall scale), so this gives the available moduli.  We have
checked this computation for the other choices of $L$ on $\bP^2$ and
find exact agreement in all cases, confirming that the Morrison--Park
model is in fact the most generic form of the Weierstrass model with
U(1) gauge group and generic charges $q = 1, 2$

We have used this Jacobian rank method to confirm in various specific cases
that the class of models defined in \cref{eq:su321weierstrass} indeed gives the
generic $\SM$ model, in the sense that the dimensionality of the moduli space
determined by the rank of the Jacobian around a random model with fixed classes
and base matches the number of moduli expected in a corresponding 6D theory
(after accounting for automorphism symmetries of the base).  We have checked
this in models of Class (A) and of Class (B). In the majority of cases, this
computation verifies that the Weierstrass model \labelcref{eq:su321weierstrass}
captures the full dimension of the moduli space. In particular, for the
``$\SU(5)$'' type B model on the base $\bP^2$, the parameters $b_2$ and $b_3$
are both in the class $-\canonclass = 3 H$ (i.e., cubics in homogeneous
coordinates on $\bP^2$).  In this case, 6D anomaly cancellation indicates that
the number of expected massless neutral hypermultiplets is  49, and the rank of
the Jacobian is 57, so we have agreement after subtracting the 8 automorphism
symmetries. However, there are cases where there is a mismatch in the number of
moduli calculated with this method and the number expected from 6D anomaly
cancellation. In \cref{sec:p2}, we discuss examples of such mismatches for
models on the base $\bP^2$.

\section{Dimensionality of models on $\bP^2$}
\label{sec:p2}

In this section, we carry out the Jacobian rank analysis described in
\cref{sec:dimension} to count the number of moduli for all 6D models described
by \cref{eq:su321weierstrass} over the base $\bP^2$. As discussed in
\cite{TaylorTurner321}, there are 98 solutions to the 6D anomaly cancellation
conditions when $T = 0$, which corresponds to F-theory on $\bP^2$. These
solutions are enumerated by choices of anomaly coefficients $b_3, b_2,
\beta$, in this case integers, which correspond in the F-theory picture to
choices of the homology classes of the corresponding parameters in the
Weierstrass model \labelcref{eq:su321weierstrass}, as described in
\cref{tab:su321homology}.

For 44 of the 98 models over the base $\bP^2$, the Jacobian rank analysis
gives a different moduli count than is expected from 6D anomaly cancellation.
There are two distinct cases: when the Jacobian rank method provides an
overcount or an undercount compared with 6D anomaly cancellation. In the
following subsections, we discuss the reason for these mismatches in both cases.

The upshot is that there are cases where there are valid anomaly cancellation
solutions, but because the anomaly coefficients are too large, the F-theory
construction discussed here develops enhanced symmetries and no longer
describes an $\SM$ model,

\subsection{When $d_2$ becomes ineffective}
\label{sec:p2-d2-ineff}

The first situation, when the Jacobian rank method provides an overcount from
the naive expectation, occurs when the gauge group enhances beyond $\SM$ due
to certain parameters in the Weierstrass model becoming ineffective. Here, we
will discuss the case where $d_2$ is ineffective but $d_1$ remains effective.
This does not cover all cases where there is gauge group enhancement, but the
analysis for other cases is similar, and so is omitted here.

If $d_2$ is ineffective, we should consider the Weierstrass model
\labelcref{eq:su321weierstrass}, but with $d_2$ set to 0:
\begin{equation}
\begin{aligned}
f &= -\frac{1}{48} \left[s_6^2 - 4 b_1 (d_0 s_5 + d_1 s_2)\right]^2 + \frac{1}{2} b_1 d_0 \left[2 b_1 \left(d_0 s_1 s_8 + d_1 s_2 s_5\right) - s_6 (s_2 s_8 + b_1 d_1 s_1)\right]\,, \\
g &= \frac{1}{864} \left[s_6^2 - 4 b_1 (d_0 s_5 + d_1 s_2)\right]^3 + \frac{1}{4} b_1^2 d_0^2 \left(s_2 s_8 - b_1 d_1 s_1\right)^2 \\
&\quad - \frac{1}{24} b_1 d_0 \left[s_6^2 - 4 b_1 (d_0 s_5 + d_1 s_2)\right] \left[2 b_1 \left(d_0 s_1 s_8 + d_1 s_2 s_5\right) - s_6 (s_2 s_8 + b_1 d_1 s_1)\right]\,.
\end{aligned}
\end{equation}

In fact, the model gains an extra generating section when $\tuning{d_2}{0}$,
indicating that the gauge algebra is enhanced to $\asu(3) \oplus \asu(2)
\oplus \au(1) \oplus \au(1)$. The Mordell--Weil group is generated by the
sections $Q$ and $T$, which have the following $[\secx : \secy : \secz]$
components:
\begin{equation}
\begin{aligned}
Q&\colon \, \left[\frac{1}{12} \left[s_6^2 - 4 b_1 (d_1 s_2 + d_0 s_5)\right] : \frac{1}{2} b_1 d_0 (b_1 d_1 s_1-s_2 s_8) : 1\right]\,, \\
T&\colon \, \left[\frac{1}{12} \left[s_6^2 + 4 b_1 (2 d_1 s_2 - d_0 s_5)\right] : -\frac{1}{2} b_1 (b_1 d_0 d_1 s_1+d_0 s_2 s_8-d_1 s_2 s_6) : 1\right]\,.
\end{aligned}
\end{equation}
The section of the original model (with the standard model gauge group) is
equal to $-(Q + T)$, where $+$ represents elliptic curve addition.

The charged matter spectrum of the model is given in \cref{tab:d2ineffspec}.
The quantity $\Delta_a^\prime$ in the table is
\begin{equation}
\Delta_a^\prime = b_1 d_1 \left[b_1 d_1 s_1^2 + s_5 (s_2 s_5 - s_1 s_6)\right] + s_8 [s_6 (s_1 s_6 - s_2 s_5) - 2 b_1 d_1 s_1 s_2] + s_2^2 s_8^2\,.
\end{equation}
Meanwhile, $I_{Q, q = 1}$ and $I_{T, q = 1}$ are the loci supporting $(\bm{1},
\bm{1})_{1, 0}$ and $(\bm{1}, \bm{1})_{0, 1}$ matter, respectively. Let us
first focus on the $I_{Q, q = 1}$ locus. This is naively given by $\locus{\secy_Q = 3
\secx_Q^2 + f \secz_Q^4 = 0}$, but we must remove contributions from the other
loci. First, we note that $\secy_Q$ and $3 \secx_Q^2 + f \secz_Q^4$ are both
proportional to $b_1 d_0$. Loci where either $b_1 = 0$ or $d_0 = 0$ support
matter charged under the nonabelian factors and should therefore be excluded
from the locus. We can therefore drop these factors from $\secy_Q$ and $3 \secx_Q^2
+ f \secz_Q^4$ when determining $I_{Q, q = 1}$. A resultant
analysis \cite{CveticKleversPiraguaMultU1} reveals
that the remaining locus contains one copy of $\locus{b_1 = s_2 = 0}$, one copy of $\locus{b_1 = s_8 = 0}$, one copy of $\locus{s_1 = s_2 = 0}$, and one copy of $\locus{d_1 = s_8 = 0}$.
Removing the contributions of these loci leads to the multiplicity listed in
the table. The multiplicity of the $I_{T, q = 1}$ locus is calculated in a
similar way. The resulting matter spectrum satisfies the 6D gauge and mixed
anomaly constraints with
\begin{equation}
 \tilde{b}_{Q Q} = -2 \canonclass - \frac{2}{3} b_3 - \frac{1}{2} d_0\,, \quad \tilde{b}_{T T} = -2 \canonclass - \frac{2}{3} b_3\,, \quad \tilde{b}_{Q T} = \canonclass + \frac{1}{3} b_3 + Y\,.
\end{equation}

\begin{table}
\centering

\[\begin{array}{c@{\qquad\quad}c@{\!\!\!\!\!\!\!\!}c}\toprule
\text{Matter}                                            & \text{Locus}                                          & \text{Multiplicity} \\ \midrule
(\bm{3}, \bm{2})_{\mathrlap{\frac{1}{6}, -\frac{1}{3}}}  & \locus{b_1 = d_0 = 0}                                 & b_3 \cdot b_2 \\
(\bm{3}, \bm{1})_{\mathrlap{-\frac{1}{3}, -\frac{1}{3}}} & \locus{b_1 = s_1 s_6^2 - s_2 s_5 s_6 + s_8 s_2^2 = 0} & b_3 \cdot (-b_3 -b_2 - 3 \canonclass + Y) \\
(\bm{3}, \bm{1})_{\mathrlap{\frac{2}{3}, -\frac{1}{3}}}  & \locus{b_1 = s_8 = 0}                                 & b_3 \cdot (-b_3 - b_2 - 3 \canonclass - Y) \\
(\bm{3}, \bm{1})_{\mathrlap{-\frac{1}{3}, \frac{2}{3}}}  & \locus{b_1 = d_1 s_6 - d_0 s_8 = 0}                   & b_3 \cdot (-b_3 - 3 \canonclass - Y) \\
(\bm{3}, \bm{1})_{\mathrlap{\frac{2}{3}, \frac{2}{3}}}   & \locus{b_1 = s_2 = 0}                                 & b_3 \cdot Y \\
(\bm{1}, \bm{2})_{\mathrlap{-\frac{1}{2}, 0}}            & \locus{d_0 = \Delta_a^\prime = 0}                     & b_2 \cdot (-2 b_2 - 2 b_3 - 6 \canonclass) \\
(\bm{1}, \bm{2})_{\mathrlap{-\frac{1}{2}, 1}}            & \locus{d_0 = d_1 = 0}                                 & b_2 \cdot (-b_3 - 2 \canonclass - Y) \\
(\bm{1}, \bm{2})_{\mathrlap{\frac{1}{2}, 1}}             & \locus{d_0 = s_2 = 0}                                 & b_2 \cdot Y \\
(\bm{1}, \bm{1})_{\mathrlap{-1, -1}}                     & \locus{s_1 = s_2 = 0}                                 & Y \cdot (-\canonclass - b_2 - b_3 + Y) \\
(\bm{1}, \bm{1})_{\mathrlap{1, -1}}                      & \locus{d_1 = s_8 = 0}                                 & (b_3 + 2 \canonclass + Y) \cdot (b_2 + b_3 + 3 \canonclass + Y) \\
(\bm{1}, \bm{1})_{\mathrlap{1, 0}}                       & I_{Q, q = 1}                                          & b_2 \cdot (b_2 + 2 b_3 + 5 \canonclass) + (b_3 + 2 \canonclass - 2 Y) \cdot (b_3 + 3 \canonclass + Y) \\
(\bm{1}, \bm{1})_{\mathrlap{0, 1}}                       & I_{T, q = 1}                                          & (b_3 + 2 \canonclass) \cdot (b_2 + b_3 + 3 \canonclass) - (b_3 + 4 \canonclass + 2 Y) \cdot Y \\
(\bm{8}, \bm{1})_{\mathrlap{0, 0}}                       & \locus{b_1 = 0}                                       & 1 + \frac{1}{2} b_3 \cdot (b_3 + \canonclass) \\
(\bm{1}, \bm{3})_{\mathrlap{0, 0}}                       & \locus{d_0 = 0}                                       & 1 + \frac{1}{2} b_2 \cdot (b_2 + \canonclass) \\ \bottomrule
\end{array}\]

\caption{Matter spectrum for the model when $\tuning{d_2}{0}$, with
multiplicities for a 6D model. The quantities $\Delta_a^\prime$, $I_{Q, q =
1}$, and $I_{T, q = 1}$ are defined in the main text.}
\label{tab:d2ineffspec}
\end{table}

Before turning to moduli counting, let us note an interesting fact about the
spectrum. In particular, suppose that we wanted to Higgs this $\asu(3) \oplus
\asu(2) \oplus \au(1) \oplus \au(1)$ algebra down to our original $\asu(3)
\oplus \asu(2) \oplus \au(1)$ model. We would need to give VEVs to matter in
the $(\bm{1}, \bm{1})_{1, -1}$ representation, and we would need at least two
hypermultiplets of this matter to satisfy the D-term constraints. From the
table, this matter is localized at $\locus{d_1 = s_8 = 0}$ and has multiplicity
\begin{equation}
[d_1] \cdot [s_8] = (b_3 + 2 \canonclass + Y) \cdot (b_2 + b_3 + 3 \canonclass + Y) = \frac{1}{4} \left([d_2] + [d_0]\right) \cdot \left([d_2] - [d_0] - 2 \canonclass\right)\,.
\end{equation}
Now assume that $d_2$ is ineffective but that $d_1$ and $s_8$ are effective.
Then, both $[d_2] + [d_0]$ and $[d_2] - [d_0] - 2 \canonclass$ must be effective,
implying that
\begin{equation}
\canonclass \le [d_2] < 0\,, \quad -[d_2] \le [d_0] \le - 2 \canonclass + [d_2]\,.
\end{equation}
As our base is $\bP^2$, there are three possible choices for $[d_2]$: $-3 H$,
$-2 H$, and $-H$, where $H$ is the hyperplane class. For each of these
choices, we can determine the possible values of $[d_0]$ and calculate $[d_1]
\cdot [s_8]$. Miraculously, whenever $[d_2]$ is ineffective, $[d_1] \cdot
[s_8]$ is never greater than $1$. In other words, if $[d_2]$ is ineffective,
there are never enough $(\bm{1}, \bm{1})_{1, -1}$ hypermultiplets to
Higgs the
$\asu(3) \oplus \asu(2) \oplus \au(1) \oplus \au(1)$ algebra down to the
original $\asu(3) \oplus \asu(2) \oplus \au(1)$, at least for a $\bP^2$ base.
This fact suggests a possible explanation for the fact that the gauge algebra
of the Weierstrass model enhances when $[d_2]$ is ineffective: we are seeing a
phenomenon similar to that observed for non-Higgsable clusters (NHCs). Just as
in the model at hand, models with NHCs do not have enough charged
hypermultiplets to break the NHC while satisfying D-term constraints. At the
level of the Weierstrass model, NHCs occur when certain parameters (the coefficients in a power series expansion of $f$ and $g$ around the relevant divisor) become ineffective, thereby forcing the Weierstrass model to obtain gauge
singularities. This is exactly the behavior observed here for $\asu(3) \oplus
\asu(2) \oplus \au(1)$. Of course, an NHC cannot be Higgsed at all, whereas
there should be ways of Higgsing $\asu(3) \oplus \asu(2) \oplus \au(1) \oplus
\au(1)$ down to alternative gauge groups when $[d_2]$ is ineffective.
Nevertheless, the automatic enhancement of $\asu(3) \oplus \asu(2) \oplus
\au(1)$ to $\asu(3) \oplus \asu(2) \oplus \au(1) \oplus \au(1)$ seems to have
a similar origin as the enhancement seen for NHCs.

\subsubsection{Moduli counting}
\label{sec:p2-d2-ineff-counting}

\Cref{tab:modulid2ineff} shows the moduli counts for the models on $\bP^2$ with
$d_2$ as the only ineffective parameter. For most of the models, the calculated
number of moduli agrees with the expectations from the gravitational anomaly
cancellation condition. However, there are two classes of models where there is
a mismatch; these classes of models are separated off in the table by
horizontal lines:
\begin{itemize}
\item When all three of $\beta = [s_1]$, $[s_5]$, and $[s_8]$ are trivial, the
calculated number of moduli is one more than the expectation from anomaly
cancellation. This seems problematic, since the Weierstrass model should have
at most as many moduli as that expected from anomaly cancellation.
\item When $s_1$ is ineffective, the calculated number of moduli is smaller
than the expectation from anomaly cancellation. In all of these cases,
$[s_5]$ is also ineffective, and $[s_8]$ is trivial.
\end{itemize}
We will see below that we can fully account for the mismatches in both of these
cases.

\begin{table}
\centering

\setlength{\arraycolsep}{6pt}
\[\begin{array}{cccccccc}\toprule
b_3 & b_2 & \beta & Y & \text{Expected Moduli} & \text{Computed Moduli} & \text{Expected Moduli for SM} \\ \midrule
1 & 1 & 6  & 5 & 73 & 73 & 72 \\
1 & 2 & 5  & 5 & 63 & 63 & 62 \\
1 & 3 & 4  & 5 & 56 & 56 & 55 \\
1 & 4 & 2  & 4 & 45 & 45 & 44 \\
2 & 1 & 4  & 4 & 54 & 54 & 53 \\
2 & 2 & 3  & 4 & 46 & 46 & 45 \\
2 & 3 & 1  & 3 & 37 & 37 & 37 \\
2 & 3 & 2  & 4 & 41 & 41 & 40 \\
3 & 1 & 2  & 3 & 41 & 41 & 40 \\
3 & 2 & 1  & 3 & 35 & 35 & 34 \\
4 & 1 & 0  & 2 & 34 & 34 & 33 \\ \midrule
1 & 5 & 0  & 3 & 41 & 42 & 40 \\
2 & 4 & 0  & 3 & 35 & 36 & 34 \\
3 & 3 & 0  & 3 & 32 & 33 & 31 \\ \midrule
4 & 5 & -6 & 0 & 44 & 43 & 43 \\
5 & 4 & -6 & 0 & 41 & 40 & 40 \\
6 & 3 & -6 & 0 & 41 & 40 & 40 \\ \bottomrule
\end{array}\]

\caption{Moduli counting for models with ineffective $[d_2]$. The column
labeled ``Expected Moduli'' gives the number of moduli for the $\asu(3)
\oplus \asu(2) \oplus \au(1) \oplus \au(1)$ model that are expected from
6D anomaly cancellation, given the global gauge group structure consistent
with the spectrum in \cref{tab:d2ineffspec}. The column labeled ``Computed
Moduli'' gives the number of moduli calculated from the rank of the Jacobian
matrix \labelcref{eq:jacobianmatrix}. The column labeled ``Expected Moduli for
SM'' gives the number of moduli that would be expected for the $\asu(3) \oplus
\asu(2) \oplus \au(1)$ model based on 6D anomaly cancellation.}
\label{tab:modulid2ineff}
\end{table}

\paragraph{When $[s_1]$, $[s_5]$, and $[s_8]$ are trivial}
The first type of mismatch, occurring in models where $[s_1]$, $[s_5]$, and
$[s_8]$ are trivial, can be explained by the presence of extra $\U(1)$ gauge
factors. This phenomenon can most easily be seen when the elliptic fiber is
written as a cubic in a $\bP^2$ ambient space:\footnote{Note that the equation
below is written in a form that explicitly assumes $d_2$ has been set to 0,
since we are currently interested in situations where $[d_2]$ is ineffective.
One can describe situations where $d_2$ is nonzero by adding an extra $b_1 d_2
w^3$ term to the left-hand side.}
\begin{equation}
b_1 v w \left(d_0 v + d_1 w\right) + u v \left(s_2 u + s_6 w\right) + u \left(s_1 u^2 + s_5 u w + s_8 w^2\right) = 0\,.
\end{equation}
Here, $[u : v : w]$ are the homogeneous coordinates of the $\bP^2$ ambient
space. Note that this equation is satisfied when $v = s_1 u^2 + s_5 u w + s_8
w^2 = 0$. For arbitrary $s_1$, $s_5$, and $s_8$, the expression $s_1 u^2 + s_5
u w + s_8 w^2$ does not factor. But if the expression happens to factor, we
can read off two new sections of the fibration, which we refer to as $A$ and
$B$. Specifically, if we let $s_1 = \alpha_1 \alpha_2$, $s_5 = \alpha_1
\beta_2 + \alpha_2 \beta_1$, and $s_8 = \beta_1 \beta_2$, the two new sections
are given by
\begin{equation}
A\colon [u : v : w] = [-\beta_1 : 0 : \alpha_1]\,, \quad B\colon [u : v : w] = [-\beta_2 : 0 : \alpha_2]\,.
\end{equation}
In Weierstrass form, these new sections are given by $[x_A : y_A : \beta_1]$ and
$[x_B : y_B : \beta_2]$, with
\begin{equation}
\begin{aligned}
x_A &= \alpha_1^2 b_1^2 d_1^2 - \frac{1}{3} b_1 \beta_1 \left[\beta_1 d_0 \left(\alpha_2 \beta_1-2 \alpha_1 \beta_2\right) + d_1 \left(3 \alpha_1 s_6 - 2 \beta_1 s_2\right)\right] + \frac{1}{12} \beta_1^2 s_6^2\,, \\
y_A &= \frac{1}{2} b_1 \Big\{2 \alpha_1^3 b_1^2 d_1^3 + \alpha_1 b_1 \beta_1 d_1 \left[\beta_1 d_0 \left(2 \alpha_1 \beta_2 - \alpha_2 \beta_1\right) + d_1 \left(2 \beta_1 s_2 - 3 \alpha_1 s_6\right)\right] \\
&\qquad\qquad + \beta_1^2 \left(\beta_1 \beta_2 d_0 - d_1 s_6\right) \left(\beta_1 s_2 - \alpha_1 s_6\right)\Big\}\,, \\
x_B &= \alpha_2^2 b_1^2 d_1^2 - \frac{1}{3} b_1 \beta_2 \left[\beta_2 d_0 \left(\alpha_1 \beta_2 - 2 \alpha_2 \beta_1\right) + d_1 \left(3 \alpha_2 s_6 - 2 \beta_2 s_2\right)\right] + \frac{1}{12} \beta_2^2 s_6^2\,, \\
y_B &= \frac{1}{2} b_1 \Big\{2 \alpha_2^3 b_1^2 d_1^3 + \alpha_2 b_1 \beta_2 d_1 \left[\beta_2 d_0 \left(2 \alpha_2 \beta_1 - \alpha_1 \beta_2\right) + d_1 \left(2 \beta_2 s_2 - 3 \alpha_2 s_6\right)\right] \\
&\qquad\qquad + \beta_2^2 \left(\beta_1 \beta_2 d_0 - d_1 s_6\right) \left(\beta_2 s_2 - \alpha_2 s_6\right)\Big\}\,.
\end{aligned}
\end{equation}
When $[s_1]$, $[s_5]$, and $[s_8]$ are trivial, $s_1 u^2 + s_5 u w + s_8 w^2 =
0$ will automatically factor, and one must take these new sections into
account. If one considers these new sections together with the sections $Q$
and $T$ from before, one finds that the Mordell--Weil rank has increased from
two to three.\footnote{One can show that $T$ can be written as a combination
of $Q$, $A$, and $B$ under elliptic curve addition, so these four sections are
not fully independent.} In turn, the gauge algebra should now be $\asu(3)
\oplus \asu(2) \oplus \au(1) \oplus \au(1) \oplus \au(1)$.

The matter spectrum of this new model is described in
\cref{tab:matters1s5s8trivial}. The number of moduli expected from the
gravitational anomaly condition, meanwhile, is listed in
\cref{tab:modulid2ineffs1s5s8trivial}, along with the results of the Jacobian
calculation. We now see that the expected number of moduli matches the number
obtained from the Jacobian calculation. Therefore, the Weierstrass model has
all of the expected moduli when $[d_2]$ is ineffective and $[s_1]$, $[s_5]$,
and $[s_8]$ are trivial.

\begin{table}
\centering

\[\begin{array}{c@{\qquad\qquad}c}\toprule
\text{Matter}                                                                       & \text{Multiplicity} \\ \midrule
(\bm{3}, \bm{2})_{\mathrlap{\frac{1}{6}, -\frac{1}{3}, -\frac{1}{3}, -\frac{1}{3}}} & -b_3 \cdot \left(b_3 + 2 \canonclass\right) \\
(\bm{3}, \bm{1})_{\mathrlap{\frac{2}{3}, \frac{2}{3}, -\frac{1}{3}, -\frac{1}{3}}} & -b_3 \cdot \canonclass \\
(\bm{3}, \bm{1})_{\mathrlap{-\frac{1}{3}, -\frac{1}{3}, \frac{2}{3}, -\frac{1}{3}}} & -b_3 \cdot \canonclass \\
(\bm{3}, \bm{1})_{\mathrlap{-\frac{1}{3}, -\frac{1}{3}, -\frac{1}{3}, \frac{2}{3}}} & -b_3 \cdot \canonclass \\
(\bm{3}, \bm{1})_{\mathrlap{-\frac{1}{3}, \frac{2}{3}, \frac{2}{3}, \frac{2}{3}}}   & -b_3 \cdot  \left(b_3 + 2 \canonclass\right) \\
(\bm{1}, \bm{2})_{\mathrlap{\frac{1}{2}, 1, 0, 0}}                                  & \canonclass \cdot \left(b_3 + 2 \canonclass\right) \\
(\bm{1}, \bm{2})_{\mathrlap{-\frac{1}{2}, 0, 1, 0}}                                 & \canonclass \cdot \left(b_3 + 2 \canonclass\right) \\
(\bm{1}, \bm{2})_{\mathrlap{-\frac{1}{2}, 0, 0, 1}}                                 & \canonclass \cdot \left(b_3 + 2 \canonclass\right) \\
(\bm{1}, \bm{2})_{\mathrlap{-\frac{1}{2}, 1, 1, 1}}                                 & \left(b_3 + \canonclass\right) \cdot \left(b_3 + 2 \canonclass\right) \\
(\bm{1}, \bm{1})_{\mathrlap{1, 0, -1, -1}}                                          & \canonclass \cdot \left(b_3 + 2 \canonclass\right) \\
(\bm{1}, \bm{1})_{\mathrlap{0, 1, 1, 0}}                                            & \canonclass \cdot \left(b_3 + 2 \canonclass\right) \\
(\bm{1}, \bm{1})_{\mathrlap{0, 1, 0, 1}}                                            & \canonclass \cdot \left(b_3 + 2 \canonclass\right) \\
(\bm{8}, \bm{1})_{\mathrlap{0, 0, 0, 0}}                                            & 1 + \frac{1}{2} b_3 \cdot \left(b_3 + \canonclass\right) \\
(\bm{1}, \bm{3})_{\mathrlap{0, 0, 0, 0}}                                            & 1 + \frac{1}{2} \left(b_3 + \canonclass\right) \cdot \left(b_3 + 2 \canonclass\right) \\ \bottomrule
\end{array}\]

\caption{Matter spectrum for the model when $[d_2]$ is ineffective and
$[s_1]$, $[s_5]$, and $[s_8]$ are trivial, with multiplicities for a 6D
model.}
\label{tab:matters1s5s8trivial}
\end{table}
\begin{table}
\centering

\setlength{\arraycolsep}{6pt}
\[\begin{array}{cccccccc}\toprule
b_3 & b_2 & \beta & Y & \text{Expected Moduli} & \text{Computed Moduli} & \text{Expected Moduli for SM} \\ \midrule
1   & 5   & 0     & 3 & 42                       & 42       & 40 \\
2   & 4   & 0     & 3 & 36                       & 36       & 34 \\
3   & 3   & 0     & 3 & 33                       & 33       & 31 \\ \bottomrule
\end{array}\]

\caption{Moduli counting for models with ineffective $[d_2]$ and trivial
$[s_1]$, $[s_5]$, and $[s_8]$. The column labeled ``Expected Moduli'' gives
the number of moduli for the $\asu(3) \oplus \asu(2) \oplus \au(1) \oplus
\au(1) \oplus \au(1)$ model that are expected from 6D anomaly cancellation,
given the global gauge group structure consistent with the spectrum in
\cref{tab:matters1s5s8trivial}. The column labeled ``Computed Moduli'' gives
the number of moduli calculated from the rank of the Jacobian matrix
\labelcref{eq:jacobianmatrix}. The column labeled ``Expected Moduli for SM''
gives the number of moduli that would be expected for the $\SM$ model based on
6D anomaly cancellation.}
\label{tab:modulid2ineffs1s5s8trivial}
\end{table}

\paragraph{When $[s_1]$ and $[s_5]$ are ineffective}
In order to understand the models with the second type of mismatch, let us
consider what happens when we set $s_1$ and $s_5$ to zero (along with $d_2$).
The Weierstrass model is now given by
\begin{equation}
\begin{aligned}
f &= -\frac{1}{48} \left[s_6^2 - 4 b_1 d_1 s_2\right]^2 - \frac{1}{2} b_1 d_0 s_6 s_2 s_8\,, \\
g &= \frac{1}{864} \left[s_6^2 - 4 b_1 d_1 s_2\right]^3 + \frac{1}{24} b_1 d_0 s_6 s_2 s_8 \left[s_6^2 - 4 b_1 d_1 s_2\right] + \frac{1}{4} b_1^2 d_0^2 s_2^2 s_8^2\,.
\end{aligned}
\end{equation}
and the discriminant is proportional to $b_1^3 d_0^2 s_2^3 s_8^2$. However,
since $[s_2]$ and $[s_8]$ are trivial for the models under consideration, the
nonabelian part of the gauge algebra is simply $\asu(3) \oplus \asu(2)$. To
find the abelian part of the gauge algebra, we must calculate the
Mordell--Weil rank of these models. Surprisingly, the sections $Q$ and $T$,
which are independent when only $d_2$ is set to 0, become related when $s_1$
and $s_5$ are also set to 0. Specifically, $T$ becomes equal to $-2 Q$, and
$Q$ becomes equal to the section originally given for the $\SM$ model.
Therefore, when $d_2$, $s_1$, and $s_5$ are all ineffective, the gauge group
is still $\SM$. The matter spectrum, too, is essentially unchanged, but many
of the original $\SM$ representations happen to have multiplicity 0 because
$[s_2]$ and $[s_8]$ are trivial. The number of moduli calculated by our
Jacobian procedure should therefore equal that expected from the gravitational
anomaly condition for the $\SM$ model, which is exactly what is observed.

\subsection{When the discriminant vanishes identically}
\label{sec:vanishing-discriminant}

The cases where the Jacobian rank method gives an undercount of the moduli
compared with the expectation from 6D anomaly cancellation are exactly the
cases where the discriminant vanishes identically for the given choice of
parameters $b_3, b_2, \beta$. These cases are listed in
\cref{tab:p2VanishingDiscriminant}, and all lie in Class (A). In such a
situation, \cref{eq:su321weierstrass} is not a valid Weierstrass model. This
can be thought of as a more severe version of the gauge algebra enhancement
observed in the previous subsection.

\begin{table}
\centering

\setlength{\arraycolsep}{6pt}
\[\begin{array}{cccccc}\toprule
b_3 & b_2 & \beta & Y & \text{Expected Moduli} & \text{Computed Moduli} \\ \midrule
1 & 2 & 7 & 7 & 90 & 18  \\
1 & 3 & 5 & 6 & 67 & 18  \\
1 & 4 & 4 & 6 & 64 & 18  \\
2 & 2 & 5 & 6 & 67 & 18  \\
2 & 3 & 3 & 5 & 49 & 18  \\
2 & 4 & 2 & 5 & 48 & 19  \\
3 & 2 & 3 & 5 & 50 & 18  \\
3 & 3 & 1 & 4 & 37 & 19  \\
3 & 4 & 0 & 4 & 38 & 10  \\
4 & 2 & 1 & 4 & 39 & 19  \\ \bottomrule
\end{array}\]

\caption{Expected number of moduli from 6D anomaly cancellation for the gauge
algebra $\sm$ along with the number of moduli counted for the Weierstrass
model \labelcref{eq:su321weierstrass} using the Jacobian rank method
(subtracting $8$ to account for the automorphisms of the base $\bP^2$), for
all $\bP^2$ models where the Jacobian rank method provides an undercount.
These are precisely the cases where the discriminant vanishes identically.
Both $\beta$ and $Y$ are given for convenience, though either is sufficient to
determine the other, given $b_3, b_2$.}
\label{tab:p2VanishingDiscriminant}
\end{table}

It is worthwhile to discuss what happens in these cases when we try to
construct a corresponding $\SMuh$ Tate model. All of the models in
\cref{tab:p2VanishingDiscriminant} are in Class (A), and so would appear to
have an unHiggsing to a Tate $\SMuh$ model, as discussed in \cref{sec:review}.
Additionally, all of these models satisfy the Tate bound
\begin{equation}
4 b_3 + 3 b_2 + 2 \beta \le -8 K = 24\,,
\end{equation}
and so we would naively expect that this choice of parameters should yield a
good Tate model, yet this clashes with the observation that the discriminant
vanishes identically in the Higgsed $\SM$ model. In these cases, the
discriminant in fact identically vanishes for the Tate model as well.

As an example, let us try to carry out a Tate tuning of $\SMuh$ over $\bP^2$
with the gauge factors supported on divisors $u, v, w$ with homology classes
$[u] = H, [v] = 2 H, [w] = 7 H$, where $H$ is the hyperplane class in $\bP^2$.
This corresponds to the first line of \cref{tab:p2VanishingDiscriminant}.
Following the Tate algorithm, we can write the model in long Weierstrass form
as
\begin{equation}
y^2 + a_1 x y z + a_3 y z^3 = x^3 + a_2 x^2 z^2 + a_4 x z^4 + a_6 z^6\,,
\end{equation}
with
\begin{equation}
a_2 = a_2^\prime u v\,, \quad a_3 = a_3^\prime u^2 v w\,, \quad a_4 = a_4^\prime u^2 v^2 w\,, \quad a_6 = a_6^\prime u^4 v^3 w^2\,.
\end{equation}
From the homology classes of $u, v, w$ and the fact that $[a_j] = -j K_{\bP^2}
= 3 j H$, we find that
\begin{equation}
[a_2^\prime] = 3 H\,, \quad [a_3^\prime] = -2 H\,, \quad [a_4^\prime] = -H\,, \quad [a_6^\prime] = -6 H\,.
\end{equation}
The parameters in the Weierstrass model must be effective to be non-vanishing,
and so we must have $\tuning{a_3}{0}, \tuning{a_4}{0}, \tuning{a_6}{0}$.
However, the remaining parameters $a_1, a_2$ have order of vanishing $0$ along
the divisor $w$ that supports the $\SU(2)$. As the $\SU(2)$ factor is supposed
to be nontrivial, the only way that this can be the case is if the
discriminant vanishes identically. Indeed, an explicit check shows that this
is the case.

\subsection{Swampland questions}
\label{sec:p2-swampland}

As seen in the previous two subsections, there are cases where there is a
valid low-energy $\SM$ supergravity model that solves the anomaly cancellation
conditions, but the corresponding F-theory construction given here exhibits
enhancement to a larger gauge group or to a singularity structure so severe as
to no longer be valid. The former case is similar to an enhancement of $\SU(2)
\to \SU(2)^2 / \Z_2$ that was observed in \cite{MorrisonTaylorCompleteness}
when the anomaly coefficient takes the almost-maximal values $b = 10, 11$ in
$T = 0$ models, while the latter case can be viewed as a more extreme version
of this phenomenon. In general, it seems that certain anomaly free models in
6D supergravity produce additional gauge factors or enhancement under known
F-theory constructions when the anomaly coefficients are too large; it would
be nice to understand this better.  In particular, as the present construction
does not actually realize the desired 6D supergravity $\SM$ model in these
cases, these models are in the ``apparent swampland'' of theories that violate
no known quantum consistency constraints and yet have no known UV completion
as a quantum gravity theory \cite{VafaSwamp}. It remains to be seen whether
these models have some other UV completion through F-theory or another
approach to string compactification.

\section{Matching to Morrison--Park form}
\label{sec:morrison-park}

The approach used in \cref{sec:construction} to construct the generic $\SM$
model we have studied in this paper relies on the previous construction in
\cite{Raghuram34} of a Weierstrass model for theories with $\U(1)$ gauge group
and $q = 3, 4$ matter, which happens to unHiggs to the desired nonabelian
model of interest.  One might wish for a more direct argument for this
construction, or to construct other models with gauge groups of the form $G =
(G_\text{NA} \times \U(1)) / \Gamma$, where $G_\text{NA}$ is a
simply-connected nonabelian group and $\Gamma$ is a discrete
subgroup, where no convenient form for a Higgsed version of the theory is
already known. In such cases one might imagine carrying out the construction
by starting with the abelian $\U(1)$ part of the gauge group and then
combining this with the desired nonabelian Kodaira singularity structure
through a method analogous to that used in \cite{MorrisonTaylorMaS}.  We leave
a systematic effort towards such constructions for future work, but it is
perhaps useful in this regard to confirm that the class of models given by
\cref{eq:su321weierstrass} is indeed a specialization of the Morrison--Park
model, as may be expected from the presence of the abelian factor and the
generic nature of the matter representations in the theory.

Specifically, we can find a map between the parameters $b$, $c_0$, $c_1$,
$c_2$ and $c_3$ of the Morrison--Park form \cite{MorrisonParkU1} and
expressions in the $\SM$ Weierstrass model. Based on the forms of $f$ and $g$
reproduced in \cref{eq:morrison-park}, one can identify the following
expressions for the parameters in the Morrison--Park form in terms of the
parameters appearing in the $\SM$ model \labelcref{eq:su321weierstrass}:
\begin{equation}
\begin{aligned}
b   &= s_2\,, \\
c_3 &= b_1 d_0 s_1 - \frac{1}{2} s_2 s_6\,, \\
c_2 &= \frac{1}{4} s_6^2 - b_1 d_0 s_5 + \frac{1}{2} b_1 d_1 s_2\,, \\
c_1 &= b_1 \left(d_0 s_8 - \frac{1}{2} d_1 s_6\right)\,, \\
c_0 &= \frac{1}{4} b_1^2 \left(d_1^2 - 4 d_0 d_2\right)\,.
\end{aligned}
\end{equation}
This confirms that our $\SM$ Weierstrass model is in fact a specialization of
the Morrison--Park form.

\section{Range of geometries for construction}
\label{sec:bases}

In this section we make some simple observations regarding the range of
possible F-theory geometries in which this construction is relevant. Elliptic
Calabi--Yau threefolds and fourfolds are characterized by the two- or
three-dimensional complex base manifold supporting the elliptic fibration. For
6D F-theory models (base surfaces), the set of possible bases is fairly well
understood, and at least for Calabi--Yau threefolds with large $h^{2, 1}$ a
reasonably representative sample is given by the range of 61,539 allowed toric
bases \cite{MorrisonTaylorToric,TaylorWangNon-toric}.  For 4D models (base
threefolds), the set of toric bases alone seems to be of order $10^{3000}$
\cite{TaylorWangMC,HalversonLongSungAlg,TaylorWangLandscape}.  It is natural
to ask how many of these bases can support the Weierstrass models given here
for $\SM$ gauge group and generic matter.

As was observed in \cite{CveticEtAlQuadrillion} for the ``$\SU(5)$'' special
case of the construction given by the Class (B) model with $b_3 = b_2 =
-\canonclass$,  the constructions given here of both Class (A) and Class (B)
can be carried out in a straightforward fashion on a weak Fano base (i.e., one
without non-Higgsable structure).  This will give a generic model with gauge
group $\SM$, as long as the divisors satisfy the necessary bounds, subject to
the additional caveat that, as observed in \cref{sec:p2}, when the divisors
are too large the gauge group may be further enhanced or the discriminant may
vanish identically.  Since only a very small fraction of allowed bases are
weak Fano, however, and almost all bases contain some effective divisors
supporting non-Higgsable gauge groups, it is of interest to inquire to what
extent the constructions described here can be realized on bases that are not
weak Fano. The primary conclusion of the limited analysis we describe
here is that both Class (A) and Class (B) models can be constructed on
at least some bases that are not weak Fano without introducing
additional exotic matter or extra gauge group factors coupled directly
to the standard model gauge group $\SM$.

For bases that are not weak Fano, there are in general divisors supporting non-Higgsable
gauge groups (this is always true for 6D theories and generally true
at the geometric level for 4D theories). To tune the generic $\SM$ Weierstrass model we have found here
on such a base, the divisors supporting the nonabelian $\SU(3)$ and $\SU(2)$
factors of the standard model must not contain or intersect any divisors
supporting non-Higgsable gauge groups; otherwise, there can either be exotic
matter charged both under the $\SM$ group and the non-Higgsable gauge factors,
or the model can develop singularities at codimension one or two associated
with vanishing orders of $(4, 6)$ in $f, g$, which go beyond the Kodaira
classification or would appear to involve a superconformal theory. As the
Hodge number $h^{1, 1}$ of the threefold or fourfold increases, the
non-Higgsable cluster structure becomes increasingly dense and it is harder to
find divisors on which the gauge group $\SM$ can be tuned. Here we focus on
those cases that are not weak Fano but where  tuning of a decoupled
standard model-like sector may still be
possible. For clarity we focus attention on 6D models in this analysis, though
similar results hold for 4D models.

We begin by considering models of Class (A).  In this case, it is
possible to choose divisors $[b_1] = b_3, [d_0] = b_2, [s_1] = \beta$
on some bases that are not weak Fano so that all three of these divisors are
disjoint from all divisors supporting a non-Higgsable gauge group.  As
long as these divisors are small enough, this allows a Tate tuning of
the unHiggsed $\SMuh$ model with no exotic matter charged under both
this gauge group and the non-Higgsable factors, so that the associated
$\SM$ model has the same feature; the non-Higgsable gauge groups in
such a situation correspond to ``hidden sectors'' that communicate
only gravitationally with the standard model--like part of the
theory. As a concrete example, let us choose the base for a 6D model
to be the Hirzebruch surface $\F_3$.  This base contains a curve $S$
of self-intersection $S \cdot S = -3$ that supports a non-Higgsable
$\SU(3)$ gauge group.  The anticanonical class of the base is
$-\canonclass = 2 S + 5 F$, where $F$ is a $\bP^1$ fiber satisfying $F
\cdot F = 0, F \cdot S = 1$.  The curve $\tilde{S} = S + 3 F$
satisfies $\tilde{S} \cdot \tilde{S} = 3, \tilde{S} \cdot S = 0$.  If
we choose $[b_3] = [b_2] = [\beta] = \tilde{S}$, it is straightforward
to check that all the parameters in the model are effective, so we get
a construction of a model with the gauge group $\SM$ over a base that
is not weak Fano, in such a way that there is no interaction with the
non-Higgsable gauge groups.

In fact, to maintain the separation of the $\SM$ part of the theory from the
non-Higgsable factors, it is generally necessary to ensure that $b_3$ and
$b_2$ are disjoint from the non-Higgsable clusters, but $\beta$ does not need
to be fixed to be in a class that is a multiple of $\tilde{S}$ to have a good
construction.  The presence of a $\U(1)$ factor can be compatible in certain
cases with a non-Higgsable gauge factor without additional jointly charged
matter even if the $\U(1)$ is related through unHiggsing to an $\SU(2)$ factor
that intersects the divisor supporting the non-Higgsable factor; one such
construction is possible when $[b_3] = [b_2] = \tilde{S}$ and $[\beta] = S + 2
F$, for example. We give an explicit example of this kind of situation below
for a Class (B) model, where the only decoupled $\SM$ models involve this kind
of structure; the analysis of Class (A) models with $[\beta]$ not an integer
multiple of $\tilde{S}$ proceeds similarly.\footnote{Note that in these
situations the anomaly coefficient $\tilde{b}$ of the associated $\U(1)$
factor, however, must still be orthogonal to the divisors supporting
non-Higgsable gauge factors for the spectrum to satisfy anomaly cancellation;
this anomaly coefficient is shifted from the expected value on a weak Fano
base by a (possibly fractional) multiple of the divisors in the non-Higgsable
clusters.}   We thus see that there  are  models realizing the gauge group
$\SM$ without exotic matter even on the Hirzebruch surface $\F_3$, and that
there can be multiple models even for a fixed  choice of $[b_3], [b_2]$ that
are disjoint from all non-Higgsable clusters.  A similar result holds for
other bases with non-Higgsable clusters, although as mentioned above the room
for such tunings becomes increasingly constricted as $h^{1, 1}(X)$ increases.
Thus, on a general base we expect that the Class (A) models are parameterized
by two divisors that are disjoint from all non-Higgsable clusters and one
divisor  that satisfies weaker constraints.  In general, the condition that a
divisor is disjoint from all non-Higgsable clusters becomes increasingly
stringent as $h^{1, 1}$ increases; we leave a systematic analysis of the
conditions under which such divisors can be used to construct models with
decoupled $\SM$ gauge group for further investigation.

Now let us consider the models of Class (B).  In this situation, things are
somewhat different.  In particular, note that the divisor classes $b_3, b_2,
\eta = -4 \canonclass - 2 b_3 - b_2$ that support the factors of the unHiggsed
Pati-Salam group $\PS$ satisfy $2 b_3 + b_2 + \eta = -4 \canonclass$. Every
surface that is not weak Fano, however, contains some curve $C$ that has
self-intersection $C \cdot C \le -3$ and supports a non-Higgsable gauge group.
The presence of the non-Higgsable gauge factor can be associated
\cite{MorrisonTaylorClusters} with the fact that $-\canonclass \cdot C < 0$,
so that $-\canonclass$ is reducible and contains $C$ as a component.  From
this we see that for a Class (B) model, one of the divisor classes $b_3, b_2,
\eta$ must satisfy $D \cdot C < 0$ and hence must contain $C$ as a component;
furthermore, after removing this component (with possible multiplicity), the
remainder of the sum of divisors $2 b_3 + b_2 + \eta - n C$ must have positive
intersection with $C$ (since $(\canonclass + C) \cdot C = -2$ for any
effective $C$ in the base of self-intersection $C \cdot C < -2$). If the
self-intersection of $C$ is sufficiently negative, this can lead to a
singularity that goes beyond the Kodaira classification. For example, if $C
\cdot C = -12$, then we cannot have $b_3 \cdot C > 0$ or there would be a
codimension two $(4, 6)$ point.  Even if the model is allowed, the resulting
construction ends up having two properties: first, the gauge group on the
non-Higgsable cluster is increased through the component of $C$ lying in $b_3,
b_2$.  Second, there is exotic matter in the Pati--Salam enhanced model that
is charged under the non-Higgsable gauge group and the $\PS$ factor(s)
associated with the divisor(s) that intersect $C$. Despite this, as long as
$b_3$ and $b_2$ are disjoint from all divisors supporting non-Higgsable
factors, it can be possible to tune the Weierstrass model for $\SM$ without
exotic matter or codimension two $(4, 6)$ points.

As an example of tuning a Class (B) model over a base with non-Higgsable gauge
factors, let us consider again the case of the base $\F_3$.  We then have $2
b_3 + b_2 + \eta = -4 \canonclass$.  We cannot choose $b_3, b_2, \eta$ to all
be divisors that do not intersect $S$, since $-\canonclass \cdot S = -1$.  We
could for example try to choose $b_3 = b_2 = \eta = -\canonclass$ (the
``$\SU(5)$'' case), but then we see that each of these divisors contains a
reducible component of $S$.  On the one hand, this means that the discriminant
now vanishes to higher order on $S$, increasing the non-Higgsable gauge group
factor on this divisor to a larger group.  On the other hand, we have $(b_3 -
S) \cdot S = 2$, so that the gauge factor $\SU(3)$ lies on a divisor
intersecting the non-Higgsable gauge factor, so there is matter jointly
charged under the $\SU(3)$ factor and the (enhanced) non-Higgsable gauge
factor.  A similar story holds for the $\SU(2)$ factor.  So this corresponds
to a situation where it is possible to tune the gauge group $\SM$ with generic
matter, but there are exotic charged matter fields coupled under this gauge
group and an enhanced non-Higgsable gauge factor.

There are nonetheless a small set of Class (B) models, however, that can be
tuned on the base $\F_3$ to get a gauge  algebra $\sm \oplus \asu(3)$ with no
matter jointly charged under the $\sm$ and $\asu(3)$. If we choose the
divisors $b_3 = n_3 \tilde{S}, b_2 = n_2 \tilde{S}$, with $n_3, n_2 \in \Z$,
the $\asu(3)$ and $\asu(2)$ factors of $\sm$ will be disjoint from the $-3$
curve $S$ supporting the non-Higgsable $\SU(3)$ factor.  The Class (B)
condition $Y = 0$ then imposes the condition $\eta = -4 \canonclass - (2 n_3 +
n_2) \tilde{S}$, which is effective when $2 n_3 + n_2 \le 6$.  Let us consider
the simplest case, where $n_3 = n_2 = 1$, so $\eta = 5 S + 11 F$.  We can then
use \cref{tab:su321homology} to determine the classes of the Weierstrass
parameters: $[b_1] = [d_0] = \tilde{S}$; $[s_1] = -F$, so $\tuning{s_1}{0}$;
$[s_2] = 0$ as usual for Class (B) models; $[d_1] = 3 S + 7 F$, so $[d_1]
\cdot S = -2$ and $[d_1]$ therefore contains a factor of $S$ as a component;
similarly, $[d_2]$ contains two factors of $S$ and $s_5, s_6, s_8$ each
contain a factor of $S$.  From this we can read off the order of vanishing of
$(f, g)$ in the Weierstrass model \labelcref{eq:su321weierstrass} on the locus
$S$ to be $(2, 2)$, associated with the non-Higgsable $\SU(3)$ factor, which
we see is not enhanced. Because there is no matter charged under this gauge
factor, indeed this gives a construction of a Class (B) model on the base
$\F_3$ with gauge group $\sm \oplus \asu(3)$ with no exotic
matter.\footnote{We have not gone into the details of the global structure of
the gauge group here; as mentioned above, the anomaly coefficient of the
$\U(1)$ factor must be shifted to be proportional to $\tilde{S}$ to satisfy
anomaly cancellation, suggesting  $\Z_3$ torsion involving the non-Higgsable
$\SU(3)$.} Similar constructions seem to work for $n_3 = 1, n_2 \le 4$ and for
$n_3 = 2, n_2 \le 2$, so there is no further constraint on $b_3, b_2$. More
generally, we expect that on any base with non-Higgsable clusters the Class
(B) models without exotic extra matter charged under the $\SM$ group will be
parameterized by two divisors ($b_3, b_2$) that are constrained to be disjoint
from all non-Higgsable clusters. The parameters $b_3, b_2$ may be subject to
further constraints that we have not tried to analyze generally here
associated with the intersection of $\eta$ with curves supporting the
non-Higgsable clusters, although there are no such further constraints in the
$\F_3$ case we have analyzed explicitly here.

We thus see that it is possible to construct generic Weierstrass models with
the tuned $\SM$ gauge group using \cref{eq:su321weierstrass} even over base
surfaces that are not weak Fano.  Such models are quite constrained,
particularly for Class (B) models, and are not possible for the ``$\SU(5)$''
case, but may give a broad class of constructions of tuned standard
model--like theories over bases with additional non-Higgsable gauge factors
that could play the role of hidden sector dark matter.  We leave a more
comprehensive analysis of this possibility to future work.

\section{Conclusions}
\label{sec:conclusions}

In this paper we have given an explicit Weierstrass formulation for a general
F-theory model with gauge group $\SM$ and generic matter content.  Matching
with the results of \cite{TaylorTurner321}, where these models were analyzed
through less direct means, we find that this Weierstrass model naturally
describes two distinct subclasses of models, one that is parameterized by
three independent divisors in the base and one parameterized by two
independent divisors.  The first class of models gives a larger range of
possibilities for tuning the standard model gauge group over a fixed base. The
second class of models, on the other hand, automatically restricts to only the
matter representations realized in the standard model.  The construction
presented here is valid in the context of 4D $\cN = 1$ supergravity theories
as well as for 6D $\cN = (1, 0)$ supergravity.  We believe that the models
described here give the most general way of constructing F-theory models with
the standard model gauge group and generic associated matter in a way that
arises from a tuning of the geometry and does not come from a unified gauge
group broken by fluxes or incorporate supersymmetrically non-Higgsable gauge
group components. These constructions present a broad generalization of the
class of models recently considered in \cite{CveticEtAlQuadrillion}, and
provide an interesting playground for considering a broad class of standard
model realizations in the context of supersymmetric F-theory
compactifications.

There are a number of obvious directions in which this work could be extended.
It is natural to try to analyze more detailed aspects of the phenomenology of
these models, starting with the fluxes and chiral matter content, for which
purpose the explicit Weierstrass formulation given here should be a useful
tool. The most general class of models produced by this construction can
naturally produce some specific types of massive exotic matter, in particular
particles with $\U(1)$ charge $q = 2$ that are uncharged under the $\SU(3)$
and $\SU(2)$ factors, $\SU(2)$ doublets with $\U(1)_Y$ charge $3 / 2$, and
right-handed quarks with $\U(1)_Y$ charge $-4 / 3$; these and the $\SMuh$
unHiggsing provide some potentially interesting new phenomenological features.
It would also be interesting to explore in more detail what subset of the
large range of twofold and threefold bases that support elliptic Calabi--Yau
threefolds and fourfolds are compatible with this construction, and the
relative frequency of such constructions in the context of 4D flux vacua.  We
have also identified some apparent ``swampland'' models in 6D where theories compatible with anomaly cancellation conditions do not arise through the
general construction presented here; it would be interesting to try to
identify further quantum consistency constraints ruling out these models or
find alternative string constructions. On a more theoretical axis, this model
represents an explicit Weierstrass realization of a rather complicated gauge
group structure with nonabelian and abelian factors and a discrete quotient.
At present there is no general framework available for constructing
Weierstrass models with arbitrary such gauge groups; we found the model in
this paper by a somewhat serendipitous happenstance. The existence of this
model suggests that there may be a more general approach that could give
insight into such constructions.

\acknowledgments
We would like to thank Mirjam Cveti\v{c}, Jim Halverson, Patrick Jefferson,
Ling Lin, David Morrison, and Yinan Wang for helpful discussions. WT and AT
are supported by DOE grant DE-SC00012567, AT is supported by the Tushar Shah
and Sara Zion fellowship, and NR is supported by NSF grant PHY-1720321. This
research was also supported in part by NSF grant PHY-1748958. WT would like to
thank the Kavli Institute for Theoretical Physics (KITP) for hospitality
during part of this work. The authors would all like to thank the
Witwatersrand (Wits) rural facility and the MIT International Science and
Technology Initiatives (MISTI) MIT--Africa--Imperial College seed fund program
for hospitality and support during the final stages of this project.

\bibliographystyle{JHEP}
\bibliography{references}
\end{document}